\pdfoutput=1

\documentclass[aps,11pt]{revtex4}
\usepackage{graphicx}
\usepackage{overpic}
\usepackage{amsmath}
\usepackage[T1]{fontenc}

\usepackage{xspace}
\newcommand{\pT}{$p_{T}$\xspace} 

\begin{document}

\title{Wavelet Analysis: Event De-noising, Shower Evolution and Jet Substructure Without Jets}

\author{J.~W.~Monk}
%\institute{Niels Bohr Institute, University of Copenhagen, Denmark}
\affiliation{Niels Bohr Institute, University of Copenhagen, Denmark}

\begin{abstract}
%\abstract{
Wavelet decomposition is a method that has been applied to signal processing in a wide range of subjects.  The decomposition isolates small scale features of a signal from large scale features, while also maintaining information about \emph{where} in the signal those features occur.  Wavelets obey particular scaling relations and are especially  suited to the analysis of systems that are self-similar and scale invariant.  They are therefore a natural tool for the study of hadron collisions.  

This paper introduces the use of wavelets for de-noising (removal of soft activity), studying the scaling behaviour of a shower, and recognising jets according to this behaviour.  This is demonstrated by processing a sample of boosted W boson Monte Carlo events together with their QCD background.  The method is quite general  and can be used as a pre-processing step in conjunction with any jet-finder or other event-shape algorithm.  The result in this simple example is a significant enhancement  in both the size and shape of the W boson mass peak, together with an improved separation of the background distribution.

\end{abstract}

\maketitle

\section{Introduction}\label{sec:introduction}

The ideas presented here grew out of attempts to identify diffractive processes at the Large Hadron Collider (LHC) using Fourier transforms \cite{Monk:2010zz}.  Intense QCD activity at the LHC masks the fragile rapidity gap signature of such processes.  A Fourier analysis decomposes a proton collision event into separate contributions to the energy flow according to their angular scale.  One drawback to Fourier transforms is that they do not give any information on \emph{where} in a signal a particular contribution occurs, and there are edge effects due to the inhermetic nature of detectors at hadron colliders.

The short-time Fourier transform (STFT) is a modification of the Fourier transform in which the sinusoidal basis functions are modified by multiplying by a Gaussian of fixed width.  This Gaussian window localises the contribution from each basis function and addresses some of the drawbacks of the basic Fourier transform.

However, the STFT is not optimal because of the fixed width of the windowing Gaussian; high frequency, small scale components may in principle be localised to a much smaller region that low frequency, large scale structures, yet the STFT uses the same width for both small- and large-scale contributions.  The mathematical tools for wavelet analysis were developed in order to overcome the limitations of the STFT.

Wavelets \cite{Daubechies:1998CP, battle} have been something of a hot topic within mathematics, with a wide range of applications and well developed sophisticated technologies.  Wavelets have had an impact in a range of diverse subjects, including geophysics, cosmology and astrophysics, speech recognition, communication and signal compression, cardiology, image processing, and analysis of financial and other systems that exhibit fractal behaviour.  Wavelet-like analysis is also employed by biological systems, including human hearing and vision.  When a physicist visually identifies a pattern of calorimeter activity on an event display as a jet, they are using wavelet-like analysis to do so.

Despite this deep utility and relevance to the natural world, wavelets have rarely been used within the field of particle physics.  Theoretical studies of scaling relations in soft QCD were made in the 1990s \cite{Greiner:1995if}, long before the advent of modern jet analysis techniques at the LHC.  More recently, wavelets were proposed as a method for uncovering faint event shapes caused by jet quenching in heavy ion collisions \cite{Dremin:2007ph}, which is not dissimilar to my original plan to study diffraction using harmonic analysis.  There has also been a recent study on wavelets and boosted bosons \cite{Rentala:2014bxa} that is coincident with the work outlined here.  Although that study ostensibly shares the same topic, the present work is unrelated in either form or origin, and takes a different approach.

On applying the wavelet transform, a signal is decomposed into a set of wavelet basis functions, which in the case of the discrete version of the transform are usually orthogonal.  This decomposition is formally expressed in equation \ref{eqn:decomposition}

\begin{equation}
P\left(\phi\right) = \sum_{m,n}C_{mn}\psi\left(2^{\left(m-M\right)}\frac{\phi}{S_{0} } - n\right)\label{eqn:decomposition}
\end{equation}
where $\psi\left(x\right)$ is the wavelet basis function and $S_0$ is an arbitrary scale choice, typically the resolution limit of the signal, that is, $S_0 = \Phi/2^{M}$, where $\Phi$ is the physical length of the signal and $M$ is the number of samples present in the discrete signal.  The index $n$ of the wavelet coefficient $C_{mn}$ in equation \ref{eqn:discreteDefn} is used as a translational parameter in the wavelet basis function $\psi$.  The index $m$ scales the basis function; the wavelet basis functions at two different scale indices $k$ and $m$ are related by

\begin{equation}
\psi\left(k, \phi\right)  = \sqrt{\frac{2^{k}}{2^{m}}} \psi\left(m, \frac{2^{k}}{2^{m}}\phi\right)
\label{eqn:waveletRescaling}
\end{equation}
in other words, the basis function at scale $k$ is simply a re-scaling of the basis function at scale $m$.  This scaling relation is lacking from the STFT and is the key feature that makes the wavelet transform an exciting tool for analysing hadron collisions.    Note also that, as with the choice of scale in any renormalisable system, the choice of $S_{0}$ is in principle arbitrary, though in practice it will be limited by the resolution of any measuring apparatus.  Wavelets are therefore a natural tool for analysing scale invariant and self similar systems.

\begin{equation}
C_{mn} = \sqrt{\frac{2^{\left(m-M\right)}}{S_{0}}} \int\bar{\psi}\left(2^{\left(m-M\right)}\frac{\phi}{S_{0} } - n\right)P\left(\phi\right)d\phi
\label{eqn:discreteDefn}
\end{equation}

The coefficients of the discrete wavelet transform, $C_{mn}$, are given by equation \ref{eqn:discreteDefn} , which is evaluated on a discrete dyadic grid, so that incrementing the scale index $m$ by one \emph{halves} the physical scale.  The translational index $n$ has integer values between $0$ and $2^{m}$, meaning that fully half of the wavelet coefficients $C_{mn}$ are used to encode the highest frequency components of the signal.  Applying a threshold to the wavelet coefficients is therefore a common method of removing high frequency noise and compressing a signal.  The dyadic grid of scales and locations reflects the uncertainty principle.  At high frequency (large $m$, small scale), there are a large number of coefficients closely spaced in $\phi$, giving good positional information, but with poor frequency information due to the large gap between frequencies.  On the other hand, at low frequencies there are a small number of coefficients that are widely spaced in $\phi$, which gives poor positional information, but with good frequency resolution.

The discrete wavelet decomposition is an iterative procedure that can be understood as the combination of a low-pass and high-pass filter.  The full signal is initially convoluted with the set of smallest scale wavelet basis functions, giving the highest frequency information in the signal.  This is the high pass filter.  The signal is then \emph{re-scaled} by convoluting with the wavelet scaling function, which has the effect of removing fine detail from the signal.  This second step is the low pass filter.  The high pass filter is then run again by convoluting the now smoothed signal with the next smallest scale wavelet basis functions, and so on until only the average of the signal remains.  As a useful pedagogical exercise, an example decomposition of a sequence of eight numbers is performed by hand in Appendix \ref{apdx:example} using the Harr wavelet.

It is illuminating to compare this iterative procedure with the basic principle by which a Monte Carlo shower algorithm works, which is itself an expression  of the renormalisation group \cite{Wilson:1993dy}.  A shower algorithm uses a fixed-order splitting kernel, which is then evaluated in a sequence ordered in some evolution parameter.  Depending on the shower Monte Carlo in question, the evolution parameter can include transverse momentum, virtuality or emission angle.  Each scale in the evolution parameter results in a different emission pattern probability, and the emission at one scale provides the input for the splitting kernel at the next scale in the sequence.  The emission pattern at one scale is nothing more than a re-scaling of the pattern at another scale.  If the shower splitting kernel were to satisfy the requirements for a wavelet basis function, then the shower evolution is exactly the same as the reconstruction of a signal from its wavelet components.  It is an interesting and tantalising question as to whether QCD could be usefully expressed in terms of such a wavelet basis.  Even in the absence of such a fully optimised basis, there are a sufficiently large number of known wavelet functions that some of them must surely provide a useful framework in which to study QCD.  

The hope then is that wavelets can provide a method to break down a hadron collision into contributions arising from small angles, large angles and everything in between.  Doing so allows one to isolate, and remove if desired, contributions arising from non-perturbative physics, to study the shower evolution and extract observables using only emissions arising at particular scales, and to search for characteristic scales present in the event, which could indicate the presence of known or unknown physical processes. 

\section{Application of a Wavelet Transform to Hadron Collisions}\label{sec:application}

The aim of this technique is to perform the wavelet transform of the radiation pattern in an event, which will provide information on the scales at which emissions are produced.  The direction of an emission is described by two coordinates, $\phi$, the azimuthal angle, and $y$ the rapidity.  Point-like particle emissions are not suitable for input to the discrete wavelet transform, therefore before it can be analysed, the collision event must be \emph{rasterised}, that is, turned into a two-dimensional array of numbers.  A  two-dimensional histogram is produced in $y-\phi$ for each individual event.  The contents of each bin of this histogram is the sum of the transverse momenta of all the particles that fall in that bin for a given event.  This two-dimensional histogram is effectively an image of the radiation emission pattern in the event, with each bin an image pixel, and the sequence of pixels will is used as the input to the discrete wavelet transform.  The size of the bins or pixels determines the limit of the angular resolution available in the wavelet transform.

Rasterising the event in this way acts as a convenient prophylactic against infra-red divergences.  While a formal proof of infra-red safety is notoriously difficult \cite{Salam:2007xv}, intuitively, any result obtained from the rasterised representation cannot be sensitive to soft or collinear emissions because such emissions do not change the transverse momentum sums of any of the image pixels.  The rasterised event is also rather similar to the structure of calorimeters used in modern particle physics detector experiments.  The size of the image grid in $\phi$ is $2\pi$, and to ensure an almost equal coverage in rapidity, it is convenient to take a rapidity range of $\left|y\right|\le3.2$, giving nearly square pixels.  Symmetry between the $\phi$ and $y$ coverage is not a requirement of the transform, but it makes the results easier to interpret because  the wavelet scaling factor in the $\phi$ direction will have the same meaning as the scaling factor in the $y$ direction.  For the same reason, it is also convenient to have an equal number of histogram bins, $N$, along the $\phi$ and $y$ axes, with $N$ being radix two a requirement of the discrete wavelet transform.  In this initial study the choice $N=128$ was used, which gives an angular resolution of approximately $\Delta R = 0.05$.  This is probably somewhat smaller than the best angular resolution currently available from hadronic calorimeters, but this choice allows the technique to be studied under a best case scenario.  Advanced experimental techniques that combine charged particle tracking and fine-grained calorimetry mean such an angular resolution limit is not unfeasible with current detectors.

Having constructed the rasterised event representation, it can then be subjected to a wavelet transformation.  Several freely available computer libraries exist for carrying out wavelet transformations, although they are often specialised to tasks such as image manipulation or geophysics.  The GNU Scientific Library \cite{gsl} (GSL) includes libraries for performing wavelet transformation, and was used in this example because it is relatively simple and general, although it has the downside of not providing a large choice of basis wavelet functions.  Future studies may benefit from the development of a wavelet library designed around the needs of collider physics, and could provide a wider range of wavelet bases, possibly including newly designed bases for QCD.

The wavelet transform returns a set of $N \times N$ coefficients.  Each coefficient can be identified with a wavelet level $m$, which specifies the scale of the coefficient, and a translation index, $n$.  Since the transformation is of a two-dimensional input signal, each coefficient has two indices for the scale and two indices for the translation, which specify the scale and translation along both the $y$ and $\phi$ axes.  If the $y$ scale and translation indices are $m,n$, and the $\phi$ scale and translation indices are $j,k$, then each pair of $m-j$ values specifies a $N \times K$ sub-array  of coefficients covering all $y$ and $\phi$ translations, with $N=2^{m}$ and  $K=2^{j}$.  Each of these sub-arrays encodes the contribution to the emission pattern at a particular angular scale.

Having thus transformed to the wavelet basis and separated the contributions to the event from different angular scales, various selection criteria may be imposed.  For example, coefficients corresponding to small (or, conversely, large) angular scale may be removed.  Given the expected correlation between the wavelet scale and the shower evolution parameter, such a filter would isolate the soft or hard contributions to the radiation pattern.  Alternatively, any wavelet coefficient with magnitude below a given threshold may be removed, which is a standard technique for de-noising a signal, and has the effect of removing small contributions, regardless of their angular size.  This would have the effect of removing pile-up and underlying event contributions from the event.

\begin{figure}[h]
\begin{center}
\hspace{-0.1\textwidth}
\includegraphics[width=1.1\textwidth]{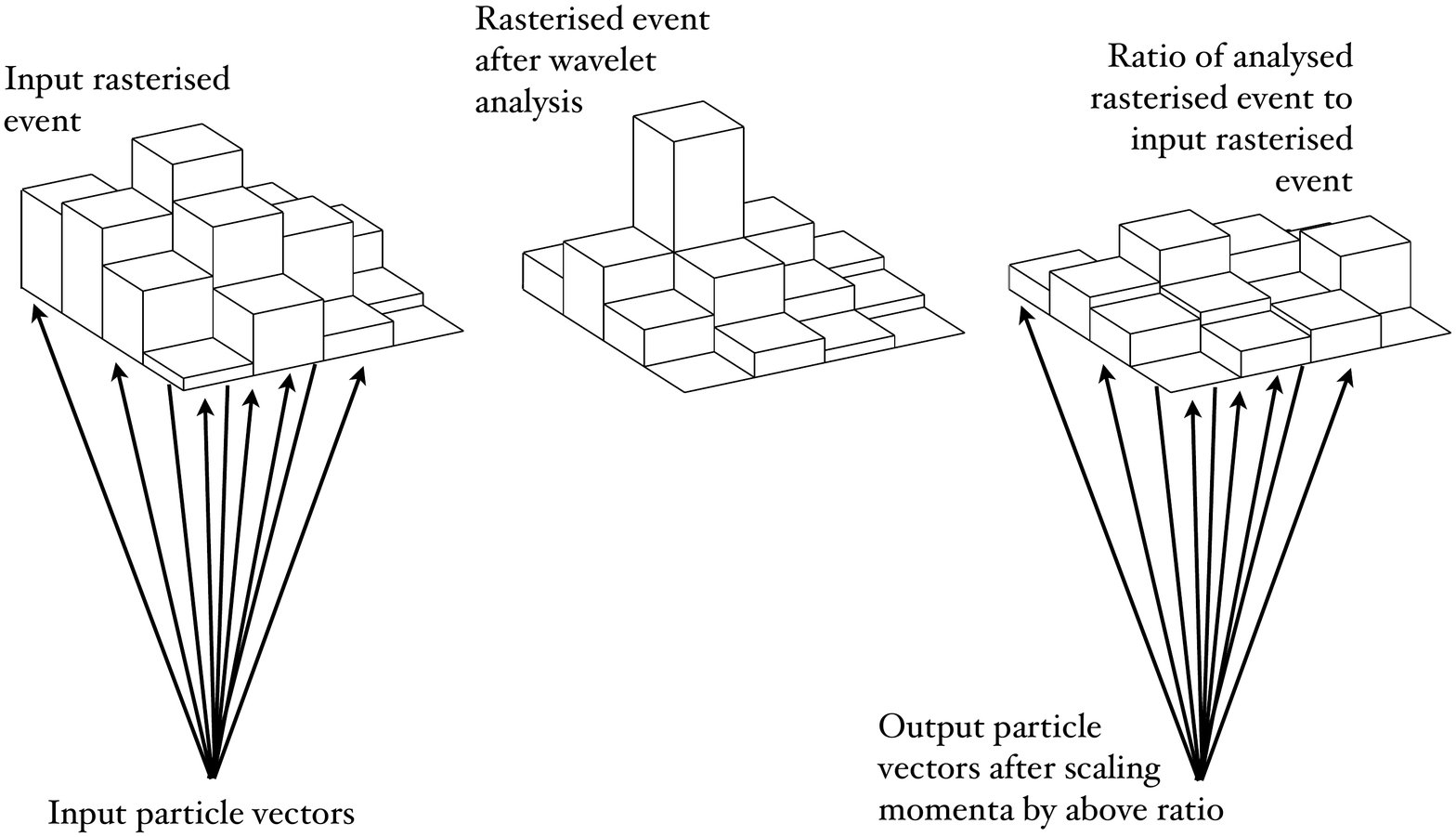}
\caption{Sequence showing the wavelet analysis of a hadron collision.  Starting with an initial set of input particle vectors, the event is first rasterised to give a two dimensional histogram (left-hand image).  Standard computational tools for discrete wavelet analysis can be applied to this rasterised representation of the event, resulting in an updated version of the rasterised event (middle image).  The ratio of the updated rasterised event to the initial rasterised event is calculated, and used to modify the input particles' momenta (right-hand image). }\label{fig:rasterisation}
\end{center}
\end{figure}

Once the desired selection and analysis is performed in the wavelet domain, the modified coefficients should be subjected to the inverse wavelet transform, which returns a now-modified version of the input rasterised event.  This $N \times N$ array is not amenable to normal analysis techniques such as jet finding, which requires particle four-vectors.  However,  the per-pixel ratio of the output rasterised event to the input rasterised event can easily be computed.    It is then possible to determine in which pixel of this ratio  each particle in the event lies, and multiply the particle momentum by that corresponding ratio.  The result is a list of particles whose momenta have been modified by the analysis and selection in the wavelet space, and which can be subjected to jet-finding or any other analysis techniques (including event shapes) in the usual way.  This process of rasterisation~$\rightarrow$~analysis~$\rightarrow$~de-rasterisation is easy to understand visually, and is sketched in figure \ref{fig:rasterisation}.

Note that the wavelet analysis can occasionally lead to negative values in the ratios; it does not makes sense to multiply a particle's momentum by a negative number, since doing so would reverse its direction of travel.  In this situation, the particle can either be removed entirely, or can be treated as a ``ghost'' particle, whose contribution can subsequently be removed from jets or other composite objects after jet-finding.

Apart from the aforementioned intrinsic suitability of wavelets for analysing QCD, this technique offers several advantages over standard jet substructure \cite{Butterworth:2008iy, Ellis:2009me} techniques:  
\begin{itemize}
\item The wavelet is applied to, and harvests information from, the entire event; whereas jet substructure operates only on jet constituents, a wavelet analysis can also use large-scale correlations that extend beyond the jet under consideration, including soft emissions at wide angles that would not normally survive jet-selection requirements.
\item The wavelet is completely independent of the jet algorithm, indeed it does not require a jet algorithm at all.  Some substructure techniques only work with certain jet algorithms, but the wavelet analysis can be used as a pre-processing step with \emph{any} jet algorithm.  Wavelets can also be used with other event shape observables.
\item The angular scale selected in a wavelet analysis is not dependent on a jet radius parameter, $R$.  This means it is possible to isolate, for example, the wide-angle contribution to a small-$R$ jet or, conversely, the small angle contribution to a large-$R$ jet.  Standard jet filtering relies on constructing small-$R$ jets within larger-$R$ jets, and the angular size of the contributions are the same as the jet size.
\item The wavelet analysis does not remove particles.  Traditional substructure techniques remove entire particles or sub-jets that are deemed to be soft in origin, but the nature of interference in QCD means that no single particle should be labelled as originating from either hard or soft interactions.  The wavelet analysis estimates a wide- and small-angle contribution to each particle, and modifies the particle accordingly in order to isolate or remove those contributions.  This approach follows the spirit of quantum mechanics.

\end{itemize}

\section{Wavelet De-Noising of Boosted W-Bosons}\label{sec:denoise}

A sample of two million Monte Carlo events containing W bosons was generated using the Pythia 8.183 event generator \cite{Sjostrand:2007gs} with tune AU2 \cite{ATLAS:2012uec} using the CTEQ6L1 \cite{Pumplin:2002vw} parton distribution function.  A second sample of ten million  QCD events was generated to provide a background.   Events were generated weighted in transverse momentum ($p_{T}$), with a minimum partonic $p_{T}$ of 180~GeV.  Pythia is a leading order event generator, and as such was not expected to describe high \pT boson production, but this analysis was mainly concerned with the behaviour of QCD showers, for which a leading order shower Monte Carlo was fast and convenient.

\begin{figure}
\begin{center}
%\begin{overpic}[width=0.7\columnwidth, angle=270]{MJ_W_Match_Filtered}
\begin{overpic}[width=0.7\columnwidth]{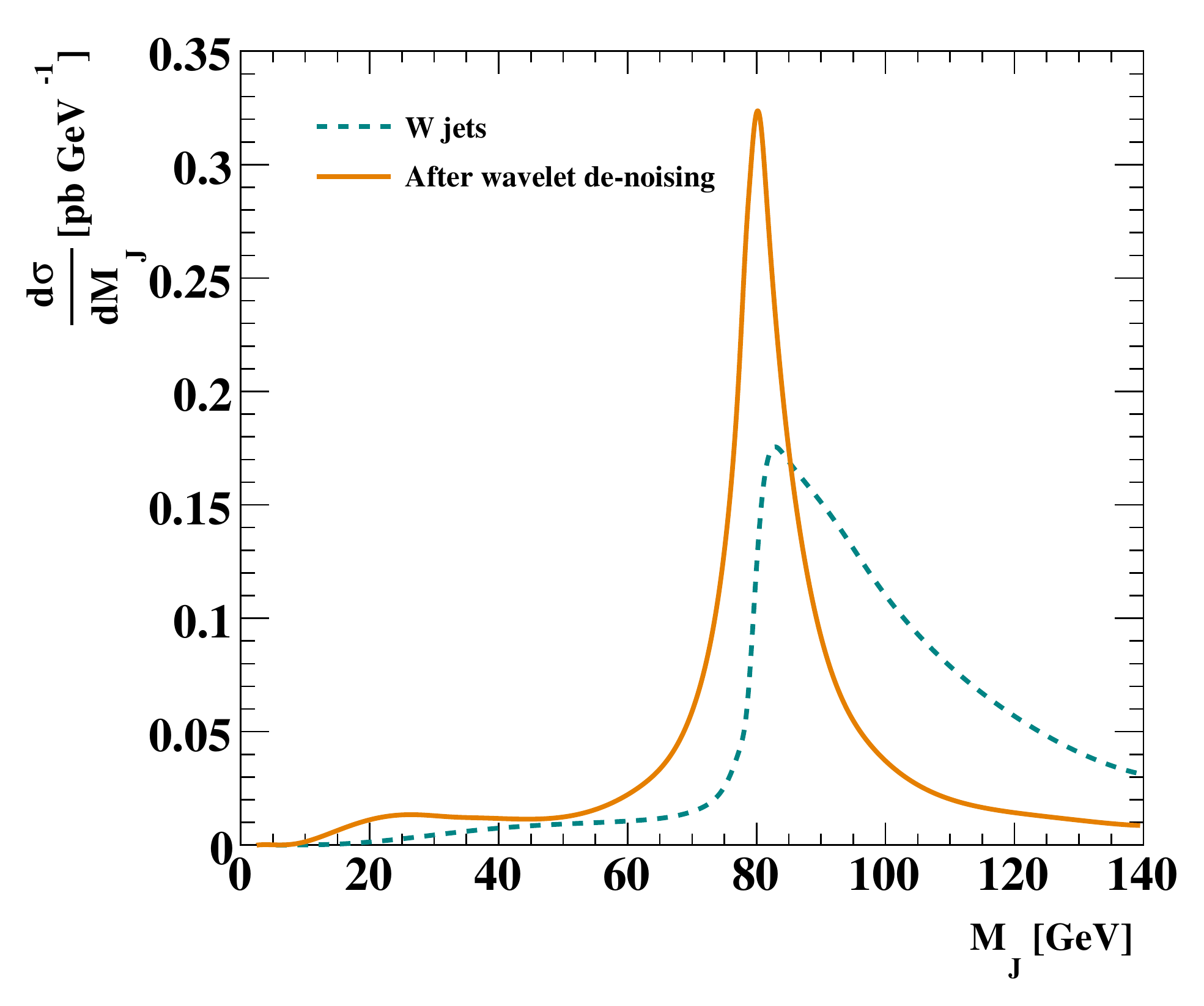}
\put(80, 70){a)}
\end{overpic}

%\begin{overpic}[width=0.7\columnwidth, angle=270]{MJ_QCD_Filtered}
\begin{overpic}[width=0.7\columnwidth]{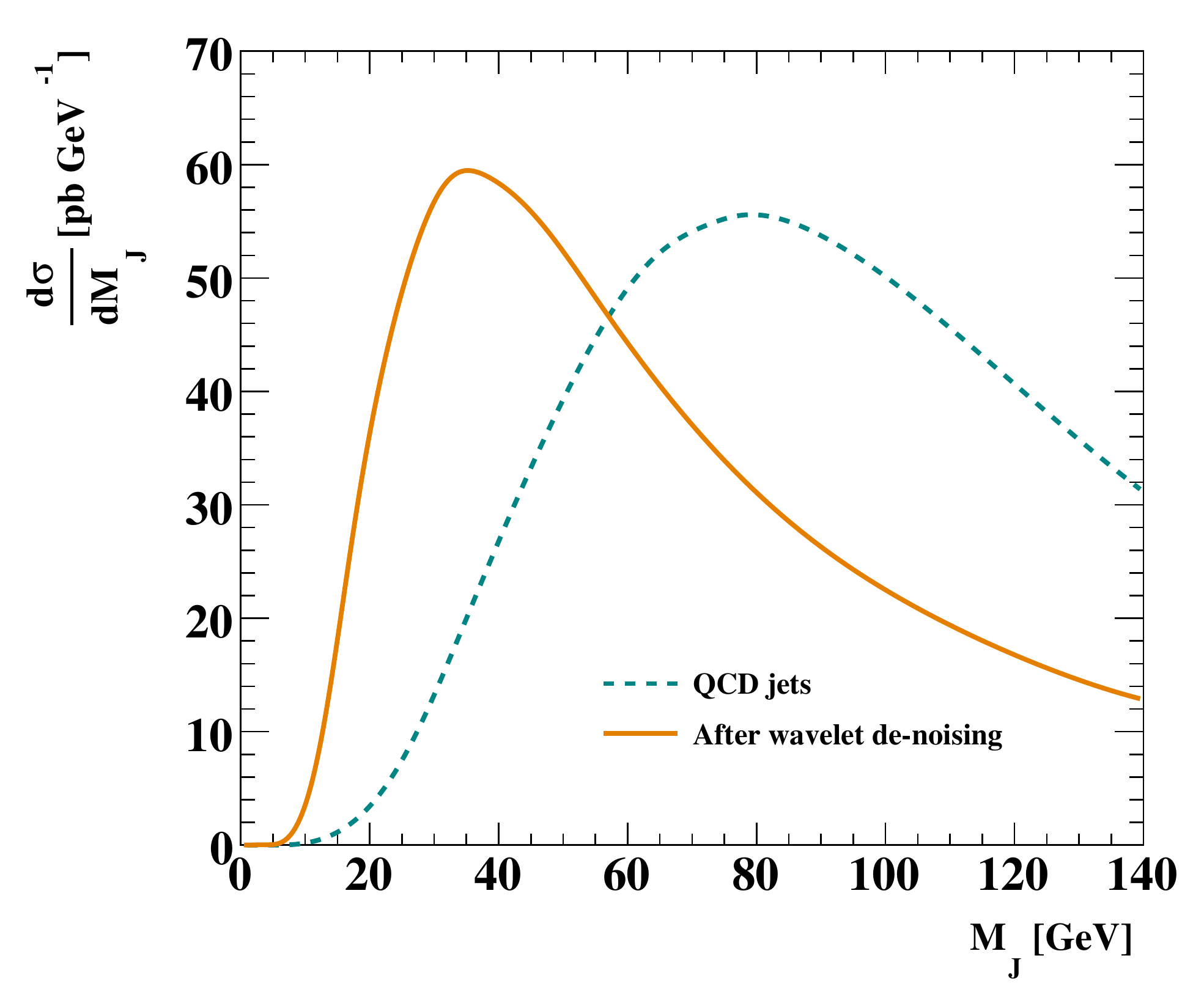}
\put(80, 70){b)}
\end{overpic}
\caption{Mass of $R=1.2$ Cambridge Aachen jets with $p_{T}\ge$~300~GeV and $\left|y\right|\le$~2 for the signal (a) and background (b) samples.  The dashed teal curve shows the mass distribution prior to wavelet filtering, the continuous orange curve shows the same distribution after performing wavelet de-noising on the collision events.}\label{fig:filteredMass}
\end{center}
\end{figure}

The samples were analysed using the Rivet framework \cite{Buckley:2010ar}, with Cambridge-Aachen jets \cite{Dokshitzer:1997in} of radius $R=1.2$ constructed using the FastJet library  \cite{Cacciari:2011ma}.  In the case of the sample of W-bosons, the jets were required to be matched to the W-boson in the event record with an angular separation no greater than $\Delta R < 1.2$.  The closest such matched jet was taken if there were multiple candidates. The masses of all such jets with \pT greater than $300$~GeV and rapidity $y\le2$ are shown in the dashed teal curves of figure \ref{fig:filteredMass}.  

Events were then subjected to the wavelet decomposition, using the Daubechies d4  wavelet and a $N\times N$ grid size of $128 \times 128$, as described in section \ref{sec:application}.  The Daubechies family of wavelets are better at encoding high frequency information than the simplest Harr wavelet. 

The resulting wavelet coefficients were filtered by setting any coefficient, $C_{jkmn}$, to zero if $\left|C_{jkmn}\right| <1$~GeV.  The noise threshold of $1$~GeV was chosen as it is approximately the scale below which QCD becomes non-peturbative.  The performance of the de-noising algorithm is not strongly dependent on the value of the noise threshold, which could in any case be optimised by studying the behaviour of soft inelastic (``minimum bias'') collisions in the wavelet domain.  

After inverting the wavelet transform and obtaining the ratio of the modified to input event, the Cambridge Aachen jet algorithm was run again, this time on the modified set of particles.  The same jet selection criteria of $p_{T} \ge 300$~GeV and $\left|y\right|\le2$ were used, and the masses of the resulting wavelet de-noised jets are shown as the orange curves of figure \ref{fig:filteredMass}.

It is immediately clear from figure \ref{fig:filteredMass} that the shape of the signal is very much improved by the simple  wavelet de-noising.  Prior to de-noising, the signal jets have significant contributions from QCD effects such as underlying event and initial and final state showers.  This gives a rather broad peak in the mass distribution, which is asymmetric and occurs somewhat above the W-boson mass of 80~GeV.  After applying the wavelet de-noising, the peak in the signal mass distribution is symmetrical, much taller and narrower, and occurs at, rather than above, the W-boson mass of 80~GeV.  

The same behaviour is observed in the background sample.  Prior to wavelet analysis, the jets in background QCD events have a broad distribution of masses with a peak just below 80~GeV, pathologically close to the W-boson mass.   After wavelet de-noising, the mass-peak in the background distribution has shifted to between 30~and~40~GeV, giving good separation with the signal peak.  The number of background jets near the W-boson mass has also been reduced by around a factor of two.

\begin{figure}[h]
\begin{center}
\hspace{-0.1\textwidth}
\includegraphics[width=0.7\textwidth]{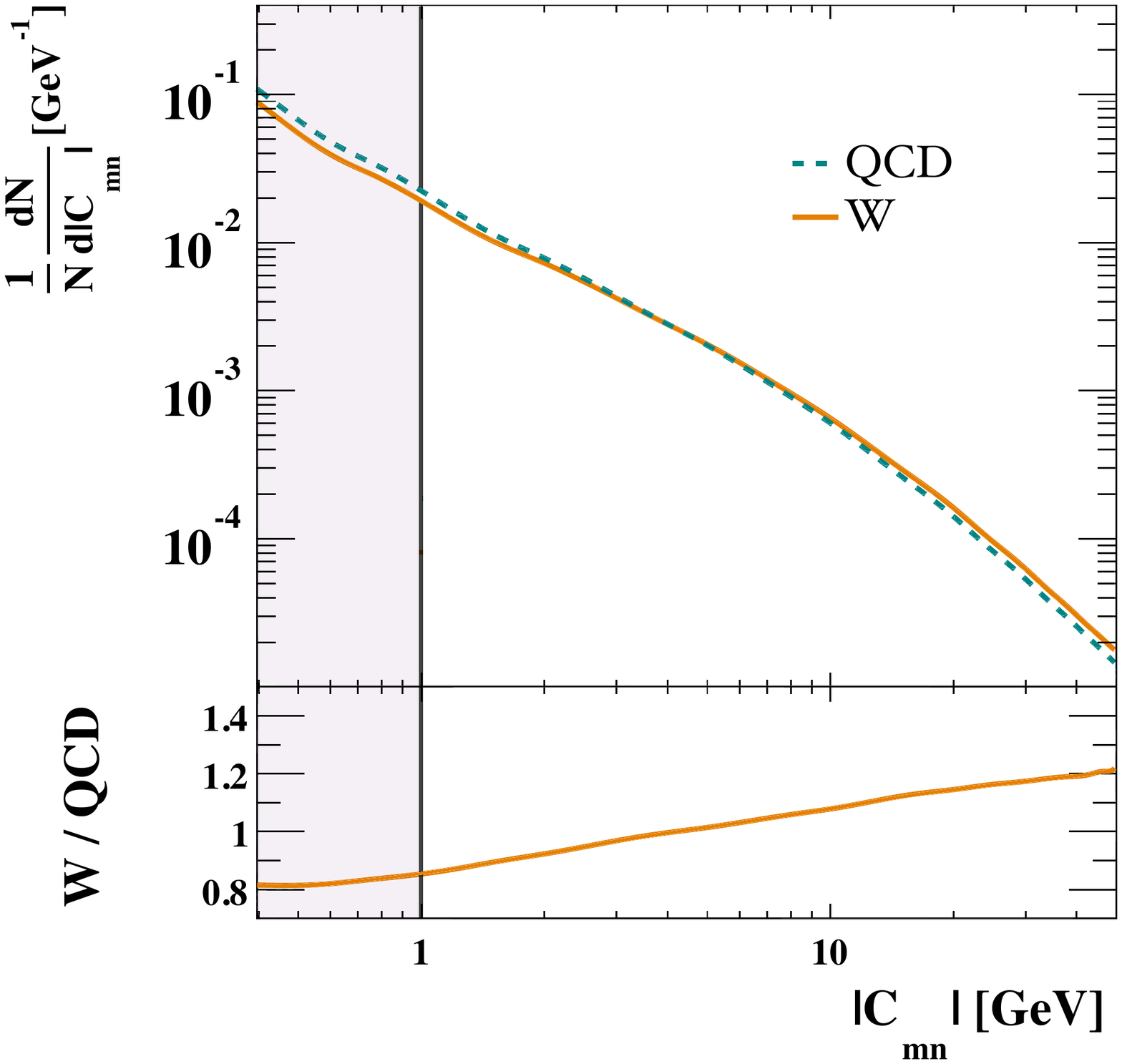}
\vspace{-0.2\textwidth}
\caption[Wavelet spectrum of QCD and W events]{Spectrum showing the magnitude of the wavelet coefficients in both QCD events (dashed teal) and W events (solid orange).  The coefficients are sharply peaked at zero, indicating that most of the activity in an event is encoded in a small number of coefficients.  The bottom panel shows the ratio of coefficients in W jet events to that in QCD jet events.  The shaded band between 0 and 1~GeV indicates the region that is removed during de-noising.}\label{fig:coeffs}
\end{center}
\end{figure}

In order to understand how it is that the de-noising process is so effective, it is useful to examine the wavelet coefficients for both W and QCD jets.  Events were selected by requiring that \emph{after} de-noising there be a jet in the event whose mass, $M_{J}$, satisfies $\left|M_{J}-80~\mathrm{GeV}\right| < 15~\mathrm{GeV}$ and, in the case of the sample of W events, that jet be matched to the W-boson.  As before, the (de-noised) jets were also required to satisfy \pT $> 300$~GeV and $\left|y\right|<$~2.  The distribution of the magnitudes of the wavelet coefficients of such events is shown in figure \ref{fig:coeffs}.  The reason why the de-noising is not particularly sensitive to the noise threshold of 1~GeV is immediately clear; the coefficients are sharply peaked near zero, and most of the activity that is removed by de-noising lies well below the threshold.  Figure \ref{fig:coeffs} also shows that, compared to QCD events, W-boson events have a slightly larger number of significant wavelet coefficients.  The number of significant wavelet coefficients is an indication of the information content of the event.  Indeed, the removal of the least significant wavelet coefficients is an effective (lossy) data compression algorithm used in a range of applications, including image and sound compression.  The larger information content of the W events indicates they posses more structure than the QCD events.

\begin{figure}
\begin{center}
%\begin{overpic}[width=0.65\columnwidth, angle=270]{rejected_y}
\begin{overpic}[width=0.65\columnwidth]{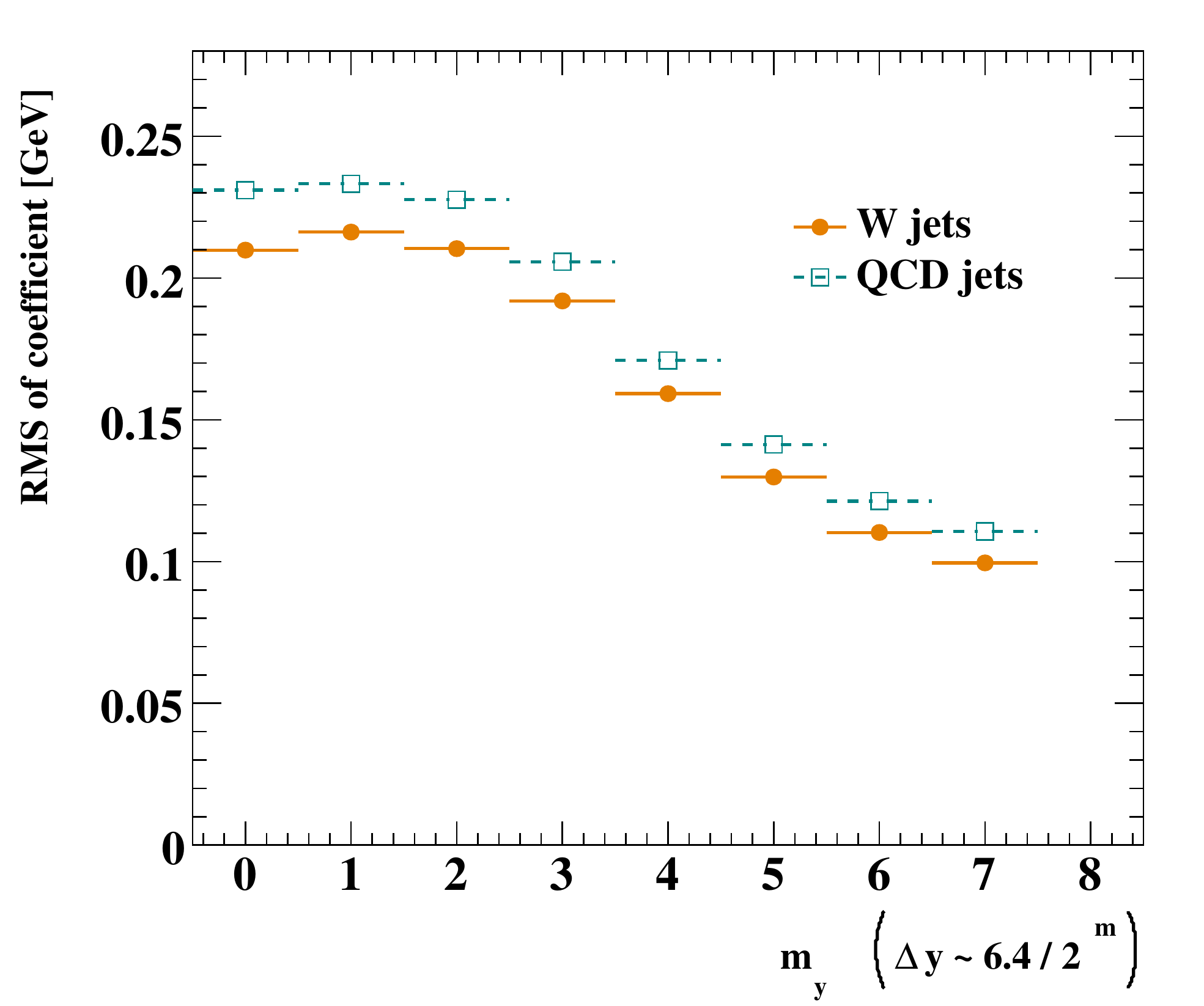}
\put(30, 25){a)}
\end{overpic}
\hspace{0.2\textwidth}
%\begin{overpic}[width=0.65\columnwidth, angle=270]{rejected_phi}
\begin{overpic}[width=0.65\columnwidth]{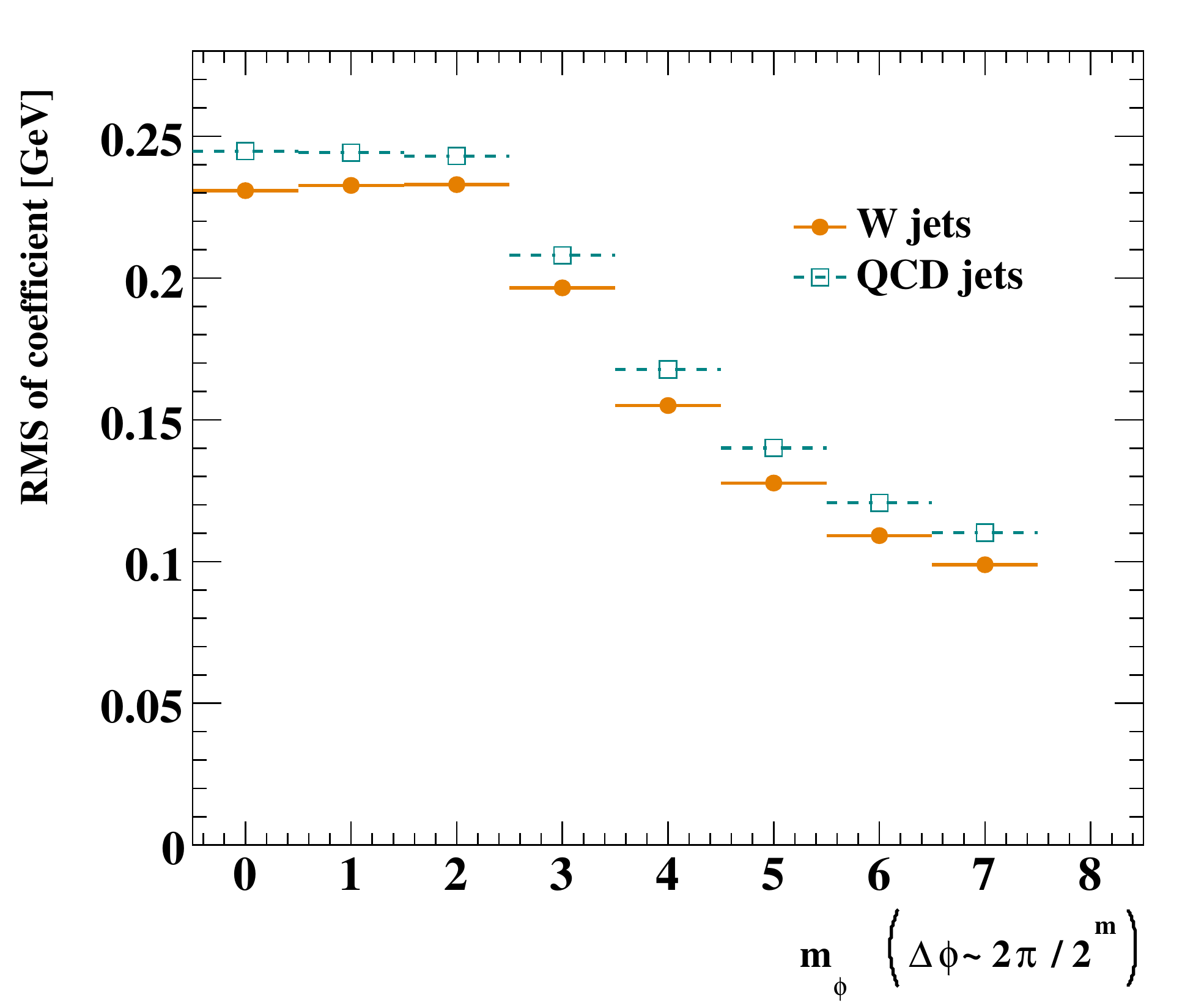}
\put(30, 25){b)}
\end{overpic}

\caption[Angular scales of activity removed by de-noising]{Angular scales of the activity removed from events by de-noising.  The x-axis displays the wavelet scale index ($m$), which indicates the angular scale of the contributions.  The y-axis shows the rms of the coefficients removed at that scale, which is an indication of the amount of activity removed.  The top plot (a) shows the rms as a function of the rapidity scale, and the bottom plot (b) shows the rms as a function of the azimuthal scale.}\label{fig:removed}

\end{center}
\end{figure}

The rms of the wavelet coefficients all having the same angular scale provides a measure of the amount of activity occurring at that scale.  The coefficients that are removed during de-noising were grouped together according to their angular scales (there are two such scales for $\phi$ and $y$), and the rms was calculated at each scale.  Figure \ref{fig:removed} shows the rms of the coefficients that are removed during de-noising as a function of the angular scales in both rapidity and azimuth.  The rms of the wide-angle coefficients that are removed is around twice that of the small-angle coefficients.  Figure \ref{fig:removed} also shows that slightly more activity is removed from QCD events compared to W-boson events, and that there is a plateau in the activity removed between scale index zero and scale index 2, which corresponds to angular size between approximately $2\pi$ and $\pi/4$.  There is also a small difference in shape between the removed activity as a function of azimuthal scale and the removed activity as a function of rapidity.  At the largest scales, the activity is flatter in azimuth compared to rapidity.  This small difference in the behaviour as a function of rapidity compared to azimuth possibly indicates the presence of colour connection effects between the beam direction and emitted partons.

\section{Wavelet based Jet Recognition}

The different behaviour of W and QCD events in the wavelet domain reflects different event structures, and creates the possibility that the two types of event may be separated by wavelet analysis.  The contribution of a given wavelet coefficient to a jet (whether de-noised or not) may be determined by rasterising the event as in figure \ref{fig:rasterisation}, setting the wavelet coefficient(s) of interest to zero and then modifying the jet constituents according to the ratio of modified-to-input rasterised event.  The difference between the unmodified jet constituents and the jet constituents after modification was considered to be the contribution to the jet from the wavelet coefficient(s) under consideration.

\begin{figure}
\begin{center}
%\begin{overpic}[width=0.65\columnwidth, angle=270]{massEvolution}
\begin{overpic}[width=0.65\columnwidth]{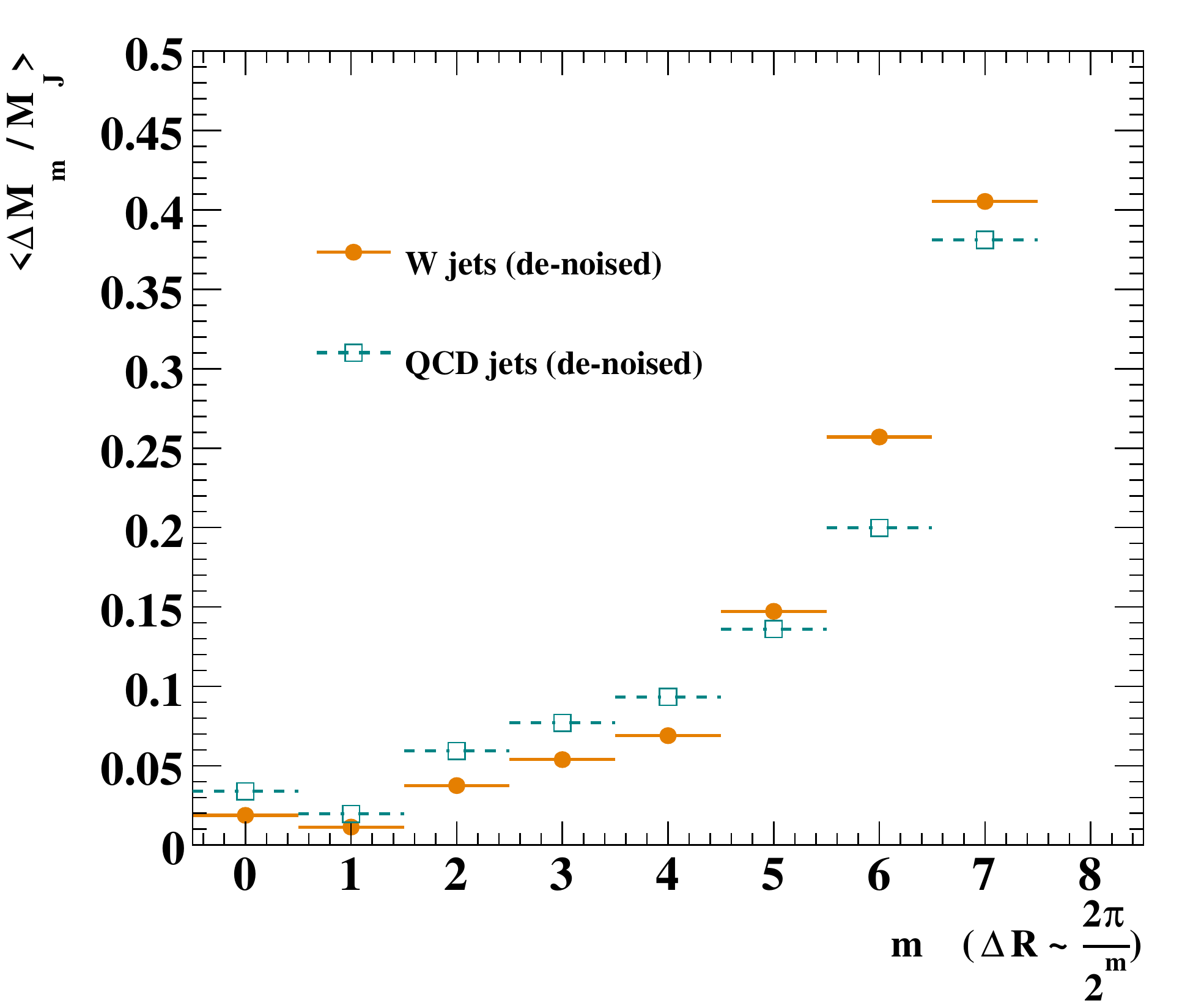}
\put(81, 26){a)}
\end{overpic}
%\begin{overpic}[width=0.65\columnwidth, angle=270]{pTEvolution}
\begin{overpic}[width=0.65\columnwidth]{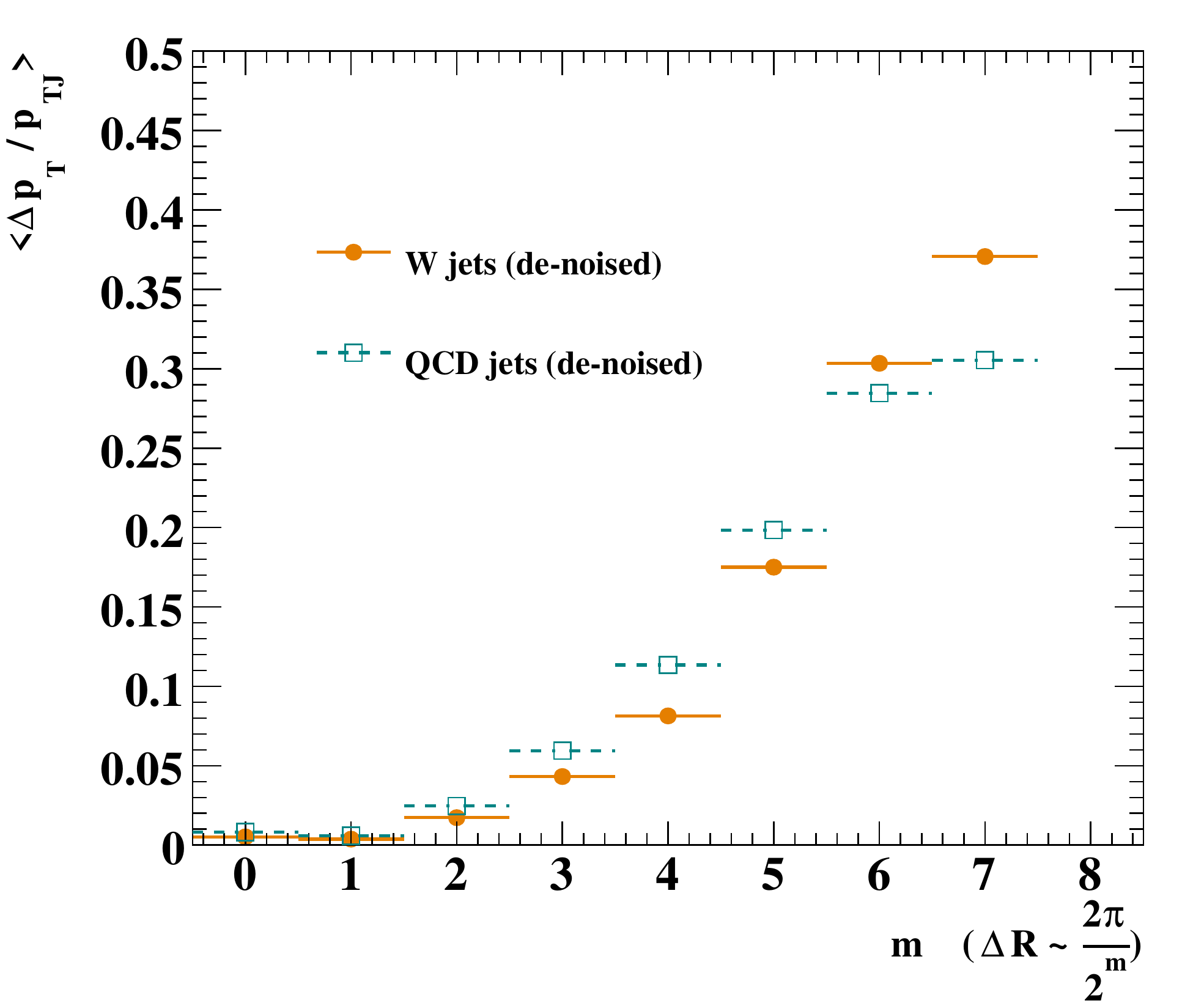}
\put(81, 26){b)}
\end{overpic}
\caption[Jet evolution profiles]{Jet evolution profiles for jet mass (a, top plot) and jet \pT (b, lower plot).  Jets have wavelet de-noising applied.  The y-axis indicates the average contribution to a jet mass or \pT arising at given angular scale, which is indicated by the wavelet level on the x-axis.  Jet evolution profiles are calculated by sequentially removing wavelet contributions, starting at the smallest scale (the largest wavelet level).}\label{fig:evolution}
\end{center}
\end{figure}

It was therefore possible to obtain a measure of how a jet evolves with angular scale.  Examples of such \emph{jet-evolution} profiles are shown in figure \ref{fig:evolution} for jet mass and jet \pT.  Only jets satisfying $\left|M_{J} - 80~\mathrm{GeV}\right| < 15\mathrm{GeV}$ were used, and in the case of the W sample, jets were required to be matched to the W-boson as described in section \ref{sec:denoise}. The y-axis in figure \ref{fig:evolution} shows the fraction of a jet's mass or \pT that is contributed by keeping only those wavelet coefficients with both rapidity and azimuthal levels below the value stated on the x-axis.  The profile was produced by initially including all but the very smallest scale contributions, in this case, those coefficients whose azimuthal and rapidity levels are both smaller than 7.  The difference between the jet mass obtained by including only those coefficients and the full jet mass is the contribution to the jet mass that occurs at wavelet level 7.  Similarly, the difference between the jet mass using only coefficients with levels smaller than 6 and the jet mass using coefficients with levels smaller than 7 is the contribution to the jet mass occurring at level 6.  The procedure was repeated to give the jet mass occurring at all levels down to level 0, which corresponds to the largest angular scale.  The jet mass contributions at each level were then divided by the full jet mass to give a mass fraction.  The same procedure was repeated with jet \pT instead of mass.

Figure \ref{fig:evolution} shows a clear difference between QCD- and W-jets in the evolution of both the jet mass and jet \pT.  Further, there is a difference in the evolution of the \pT and the mass.  In both cases, the W-jets have larger contributions coming from small angular-scales (large wavelet level).  At wavelet levels below around 5, which corresponds to angular scales larger than around 0.2, the QCD-jets show larger contributions than the W-jets.  The jet-mass evolution shows a larger contribution to both QCD- and W-jet masses at wavelet level zero, compared to jet \pT, which has a negligible contribution at level zero.  Level zero is the widest angular contribution, covering the full angular range.  The contribution to jet \pT grows more linearly with wavelet level than the contribution to jet-mass, although in the case of QCD-jets there is a flattening of the curve between levels 6 and 7.

\begin{figure}
\begin{center}
%\vspace{-1.9cm}
%\includegraphics[width=0.4\columnwidth, angle=270]{massFrac_0}
%\includegraphics[width=0.4\columnwidth, angle=270]{massFrac_1}
%\includegraphics[width=0.4\columnwidth, angle=270]{massFrac_2}
%\includegraphics[width=0.4\columnwidth, angle=270]{massFrac_3}
%\includegraphics[width=0.4\columnwidth, angle=270]{massFrac_4}
%\includegraphics[width=0.4\columnwidth, angle=270]{massFrac_5}
%\includegraphics[width=0.4\columnwidth, angle=270]{massFrac_6}
%\includegraphics[width=0.4\columnwidth, angle=270]{massFrac_7}

\includegraphics[width=0.4\columnwidth]{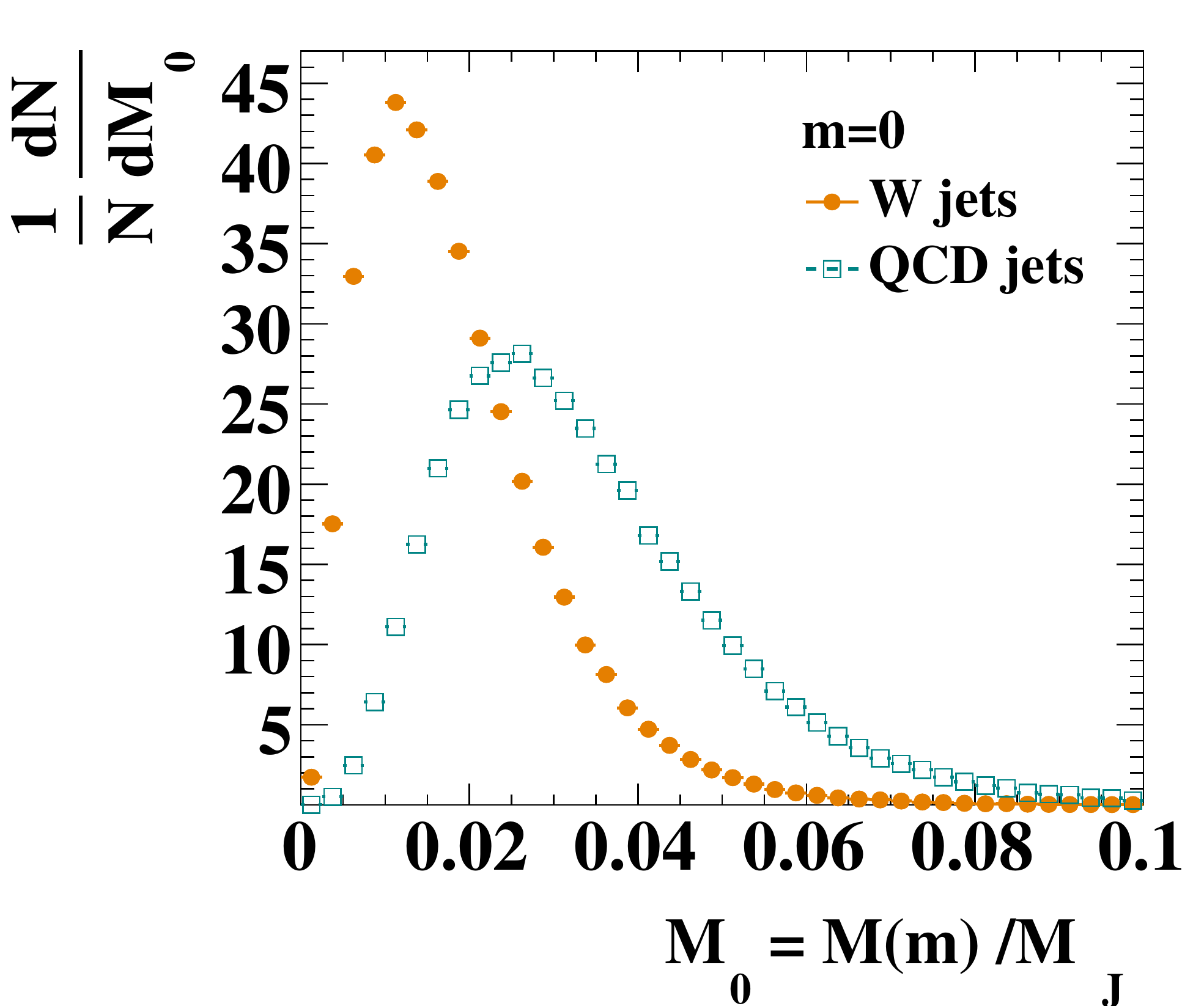}
\includegraphics[width=0.4\columnwidth]{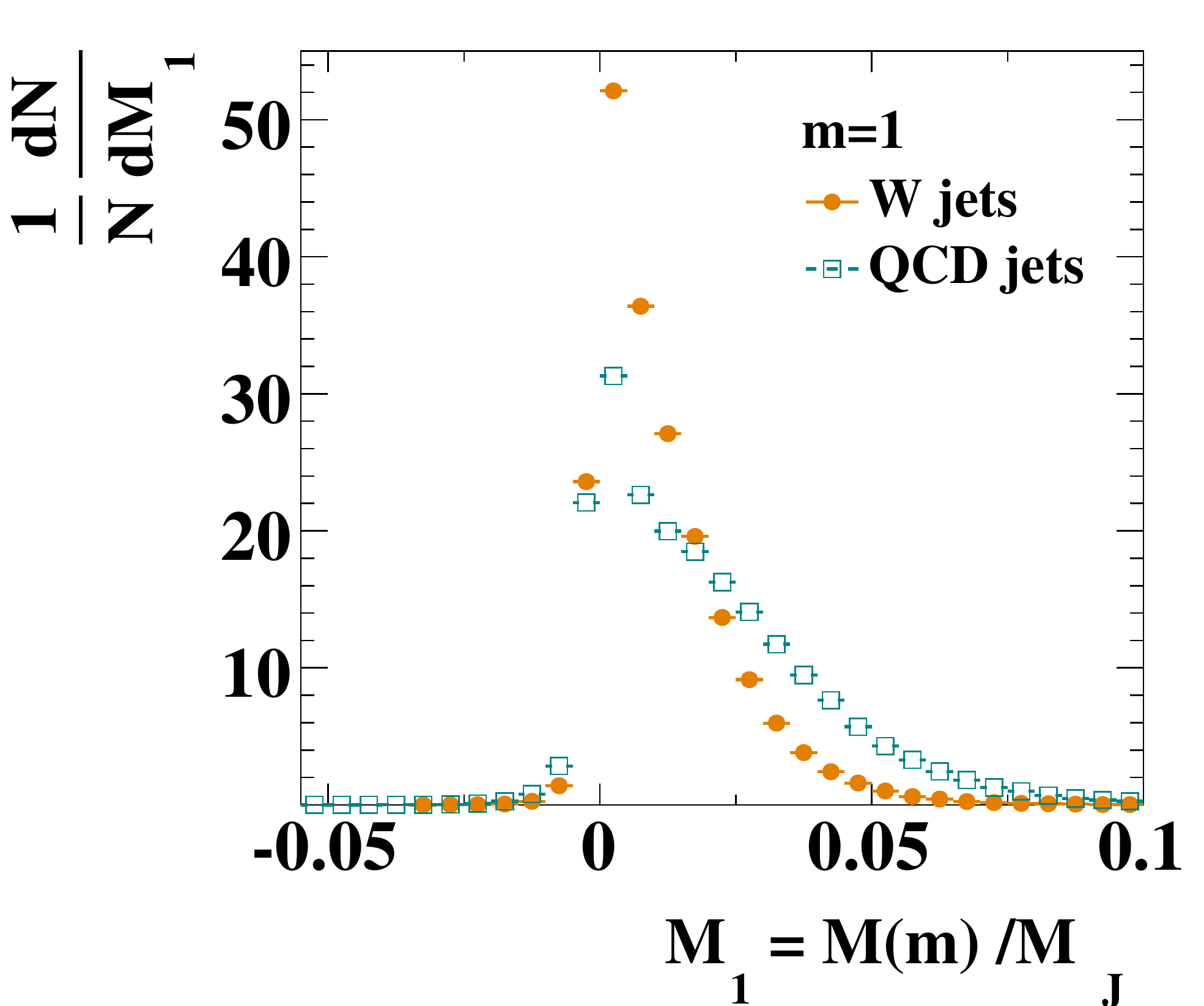}
\includegraphics[width=0.4\columnwidth]{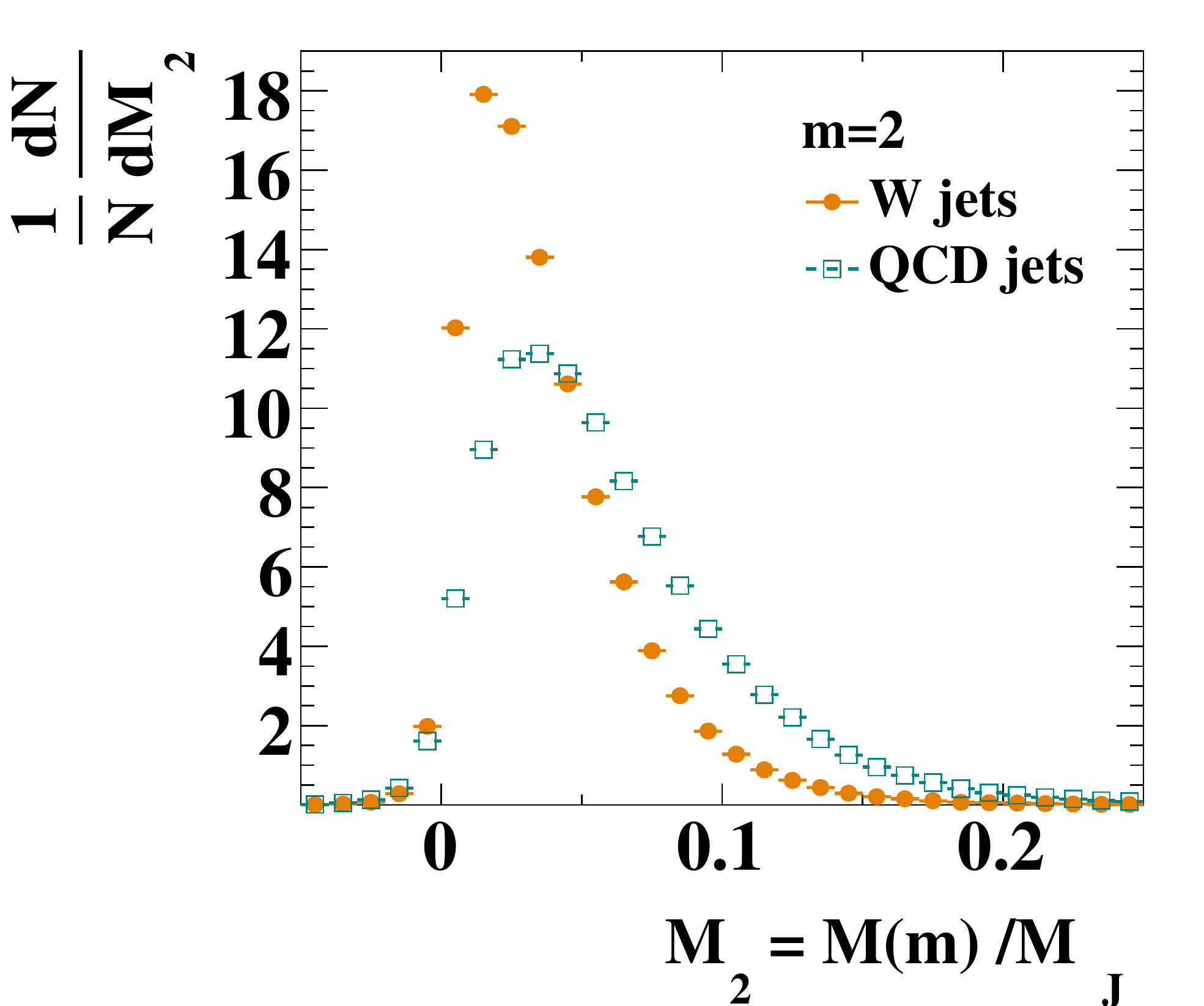}
\includegraphics[width=0.4\columnwidth]{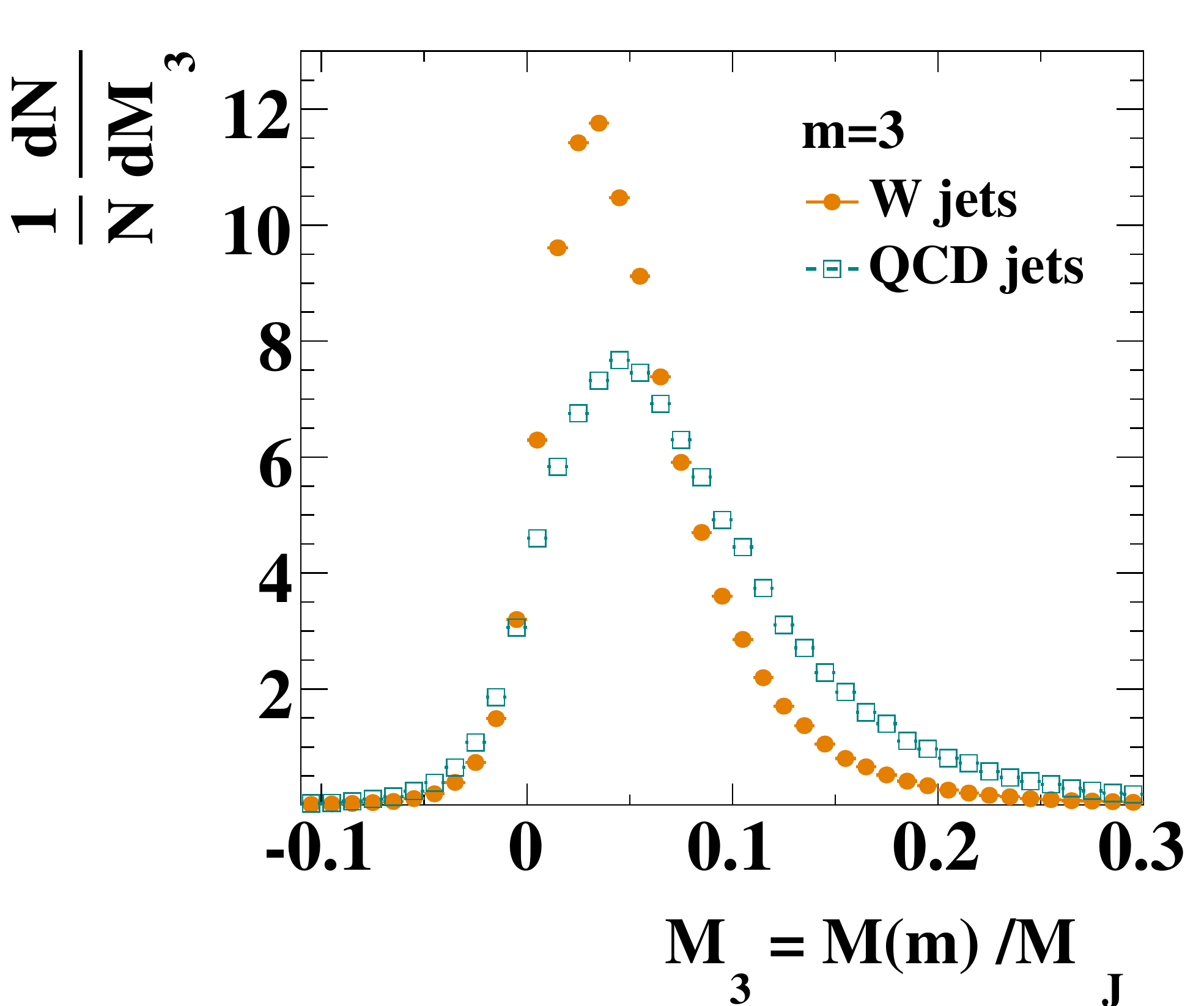}
\includegraphics[width=0.4\columnwidth]{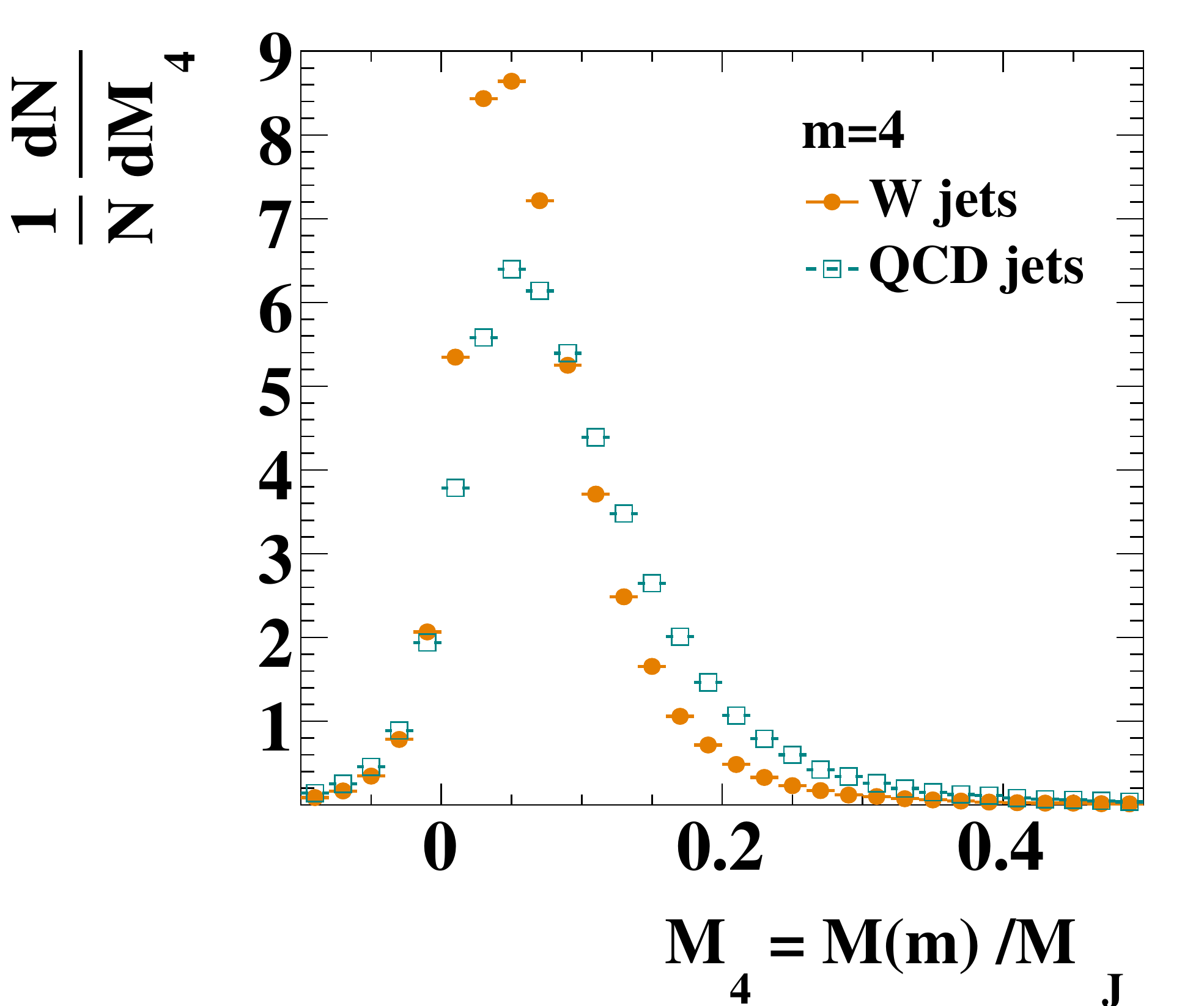}
\includegraphics[width=0.4\columnwidth]{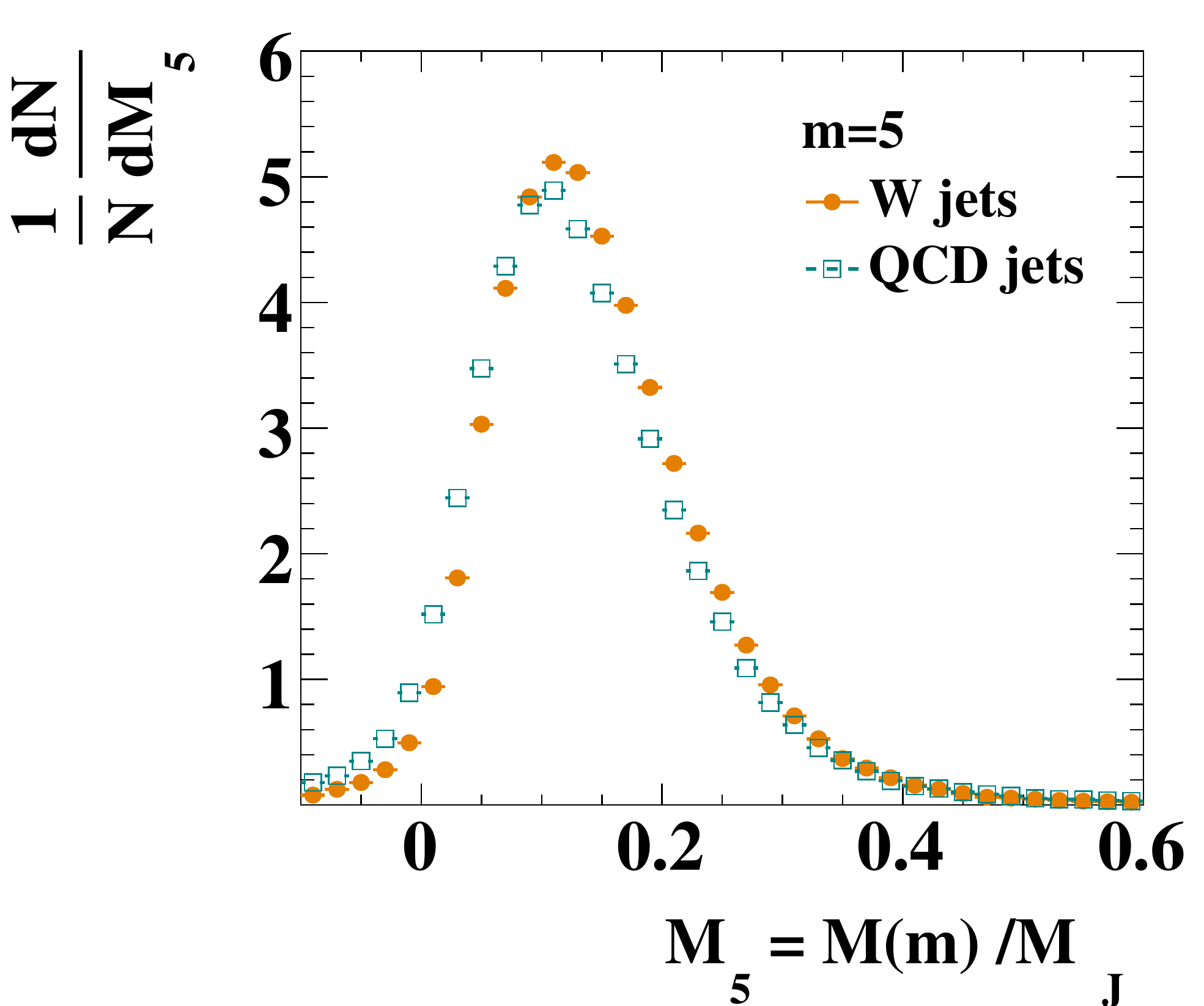}
\includegraphics[width=0.4\columnwidth]{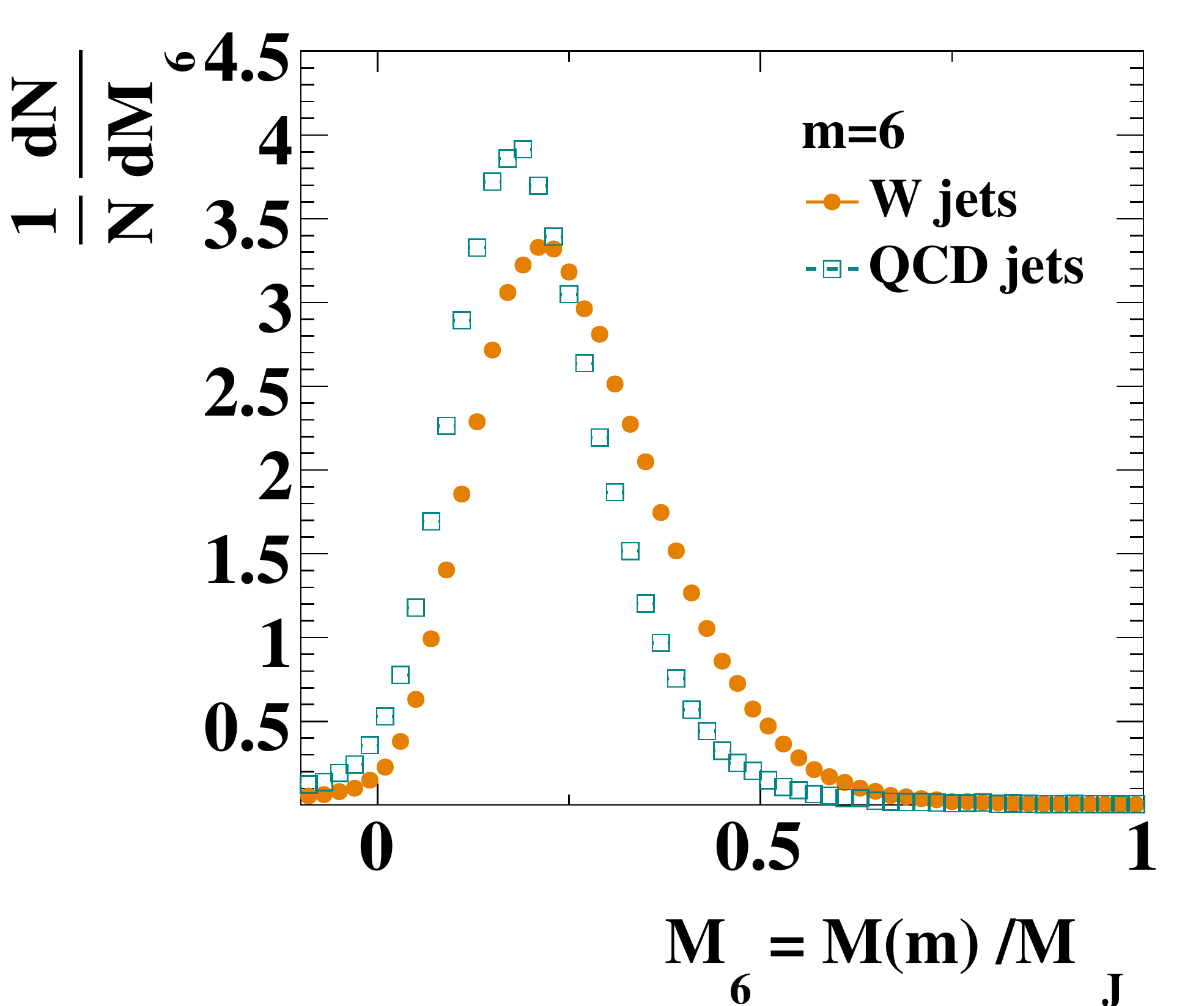}
\includegraphics[width=0.4\columnwidth]{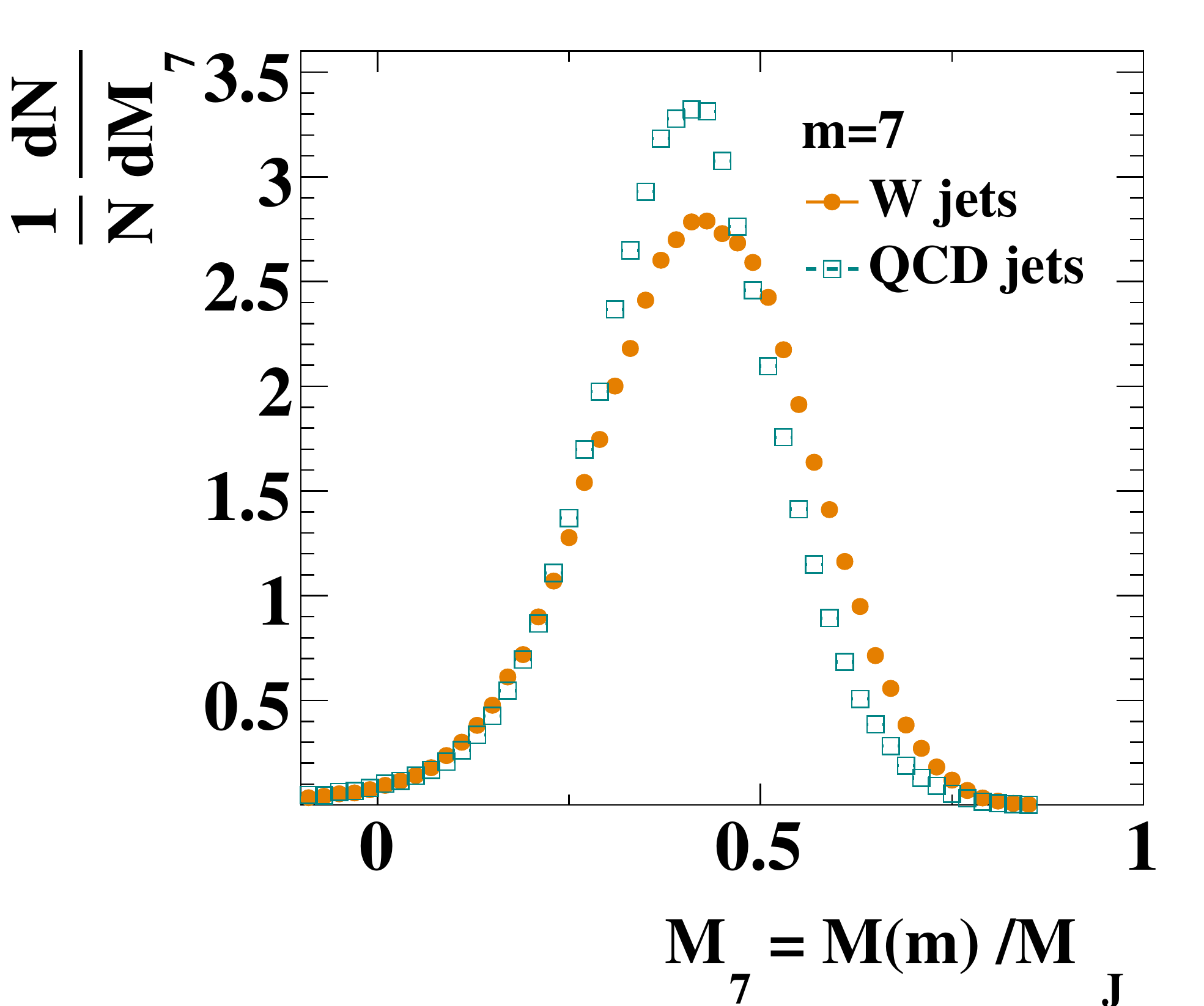}

\caption[Mass fractions]{Fractions of the jet-mass occurring at the different wavelet levels for W- and QCD-jets.  Each sub-plot corresponds to a single bin of the profile in figure \ref{fig:evolution}a}\label{fig:massFractions}
\end{center}
\end{figure}

Figure \ref{fig:massFractions} shows the de-noised jet mass fraction distributions for each of the wavelet levels of the evolution profile of figure \ref{fig:evolution}.  The difference between QCD-jets and W-jets is clearer here.  At levels below $m=5$ the W-jets have a more sharply peaked mass fraction distribution, with the peak occurring at a lower mass fraction compared to QCD-jets.  The mass fraction distribution at level $5$ is rather similar between QCD- and W-jets, and at higher wavelet levels the W-jets favour large mass fractions compared to QCD-jets.  The general trend is for W-jets to show more small-scale structure compared to QCD-jets, which tend to show somewhat more large-angle activity.  This fits the expectation for QCD-jets to be more diffuse and lack such a hard-core compared to W-jets.

The mass fraction plots of figure \ref{fig:massFractions} form a set of probability density functions (PDFs) that were used to classify the jets.  The TMVA package \cite{TMVA2007} was used to convert the eight mass fraction plots into two sets of eight PDFs, $\mathrm{P_{W}}\left(m, M_{m}\right)$ and $\mathrm{P_{QCD}}\left(m, M_{m}\right)$ for the W- and QCD-jets respectively, where $m$ is the wavelet level and $M_{m}$ is the mass fraction at that level.   The likelihood that a jet was either a W-jet or a QCD-jet was then determined by calculating its eight mass fractions at the eight wavelet levels, evaluating the PDFs and taking their product, as in equation \ref{eqn:likelihood}.

\begin{align}
L_{\mathrm{W}} &= \prod_{m=0}^{7}\mathrm{P_{W}}\left(m, M_{m}\right)\nonumber \\
L_{\mathrm{QCD}} &=  \prod_{m=0}^{7}\mathrm{P_{QCD}}\left(m, M_{m}\right)\label{eqn:likelihood}
\end{align}

\begin{figure}
\begin{center}
\includegraphics[width=0.7\columnwidth]{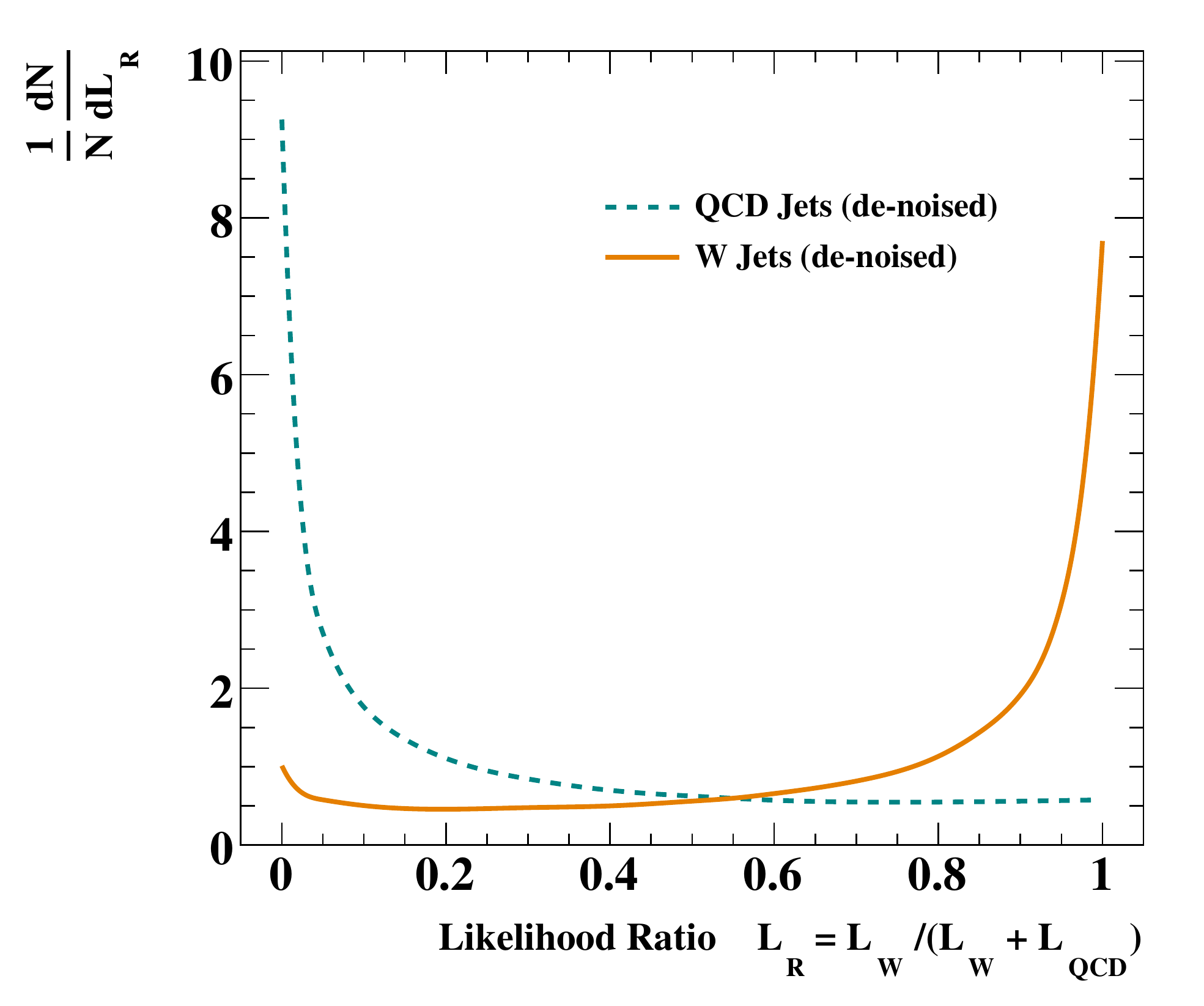}
\caption[Likelihood ratio]{Likelihood ratio for W-jets (orange) and QCD-jets (teal dashed), obtained by using the eight plots of figure \ref{fig:massFractions} as probability density functions.}\label{fig:likelihood}
\end{center}
\end{figure}

The likelihood ratio, $L_{R}$, is a measure of whether a jet is more likely to be a W-jet than a QCD-jet, and is given in equation \ref{eqn:likelihoodRatio}

\begin{equation}
L_{\mathrm{R}} = \frac{L_{\mathrm{W}}}{L_{\mathrm{W}} + L_{\mathrm{QCD}}}\label{eqn:likelihoodRatio}
\end{equation}
The $L_{\mathrm{R}}$ distribution for both de-noised W-jets and de-noised QCD-jets is shown in figure \ref{fig:likelihood}, with jets again required to have masses in the W mass window of $\left|M_{J} - \mathrm{80~GeV}\right| < \mathrm{15~GeV}$, and in the case of the sample of W-jets, be matched to a W-boson.  The QCD- and W-jets are well separated by the likelihood ratio, with W-jets strongly favouring $L_{\mathrm{R}}$ values close to 1, while QCD-jets favour values close to 0.  Selecting jets according to their $L_{\mathrm{R}}$ value is therefore a good way of rejecting QCD background jets while keeping most signal W-jets.

\begin{figure}
\begin{center}
%\begin{overpic}[width=0.7\columnwidth, angle=270]{massSelected_06}
\begin{overpic}[width=0.7\columnwidth]{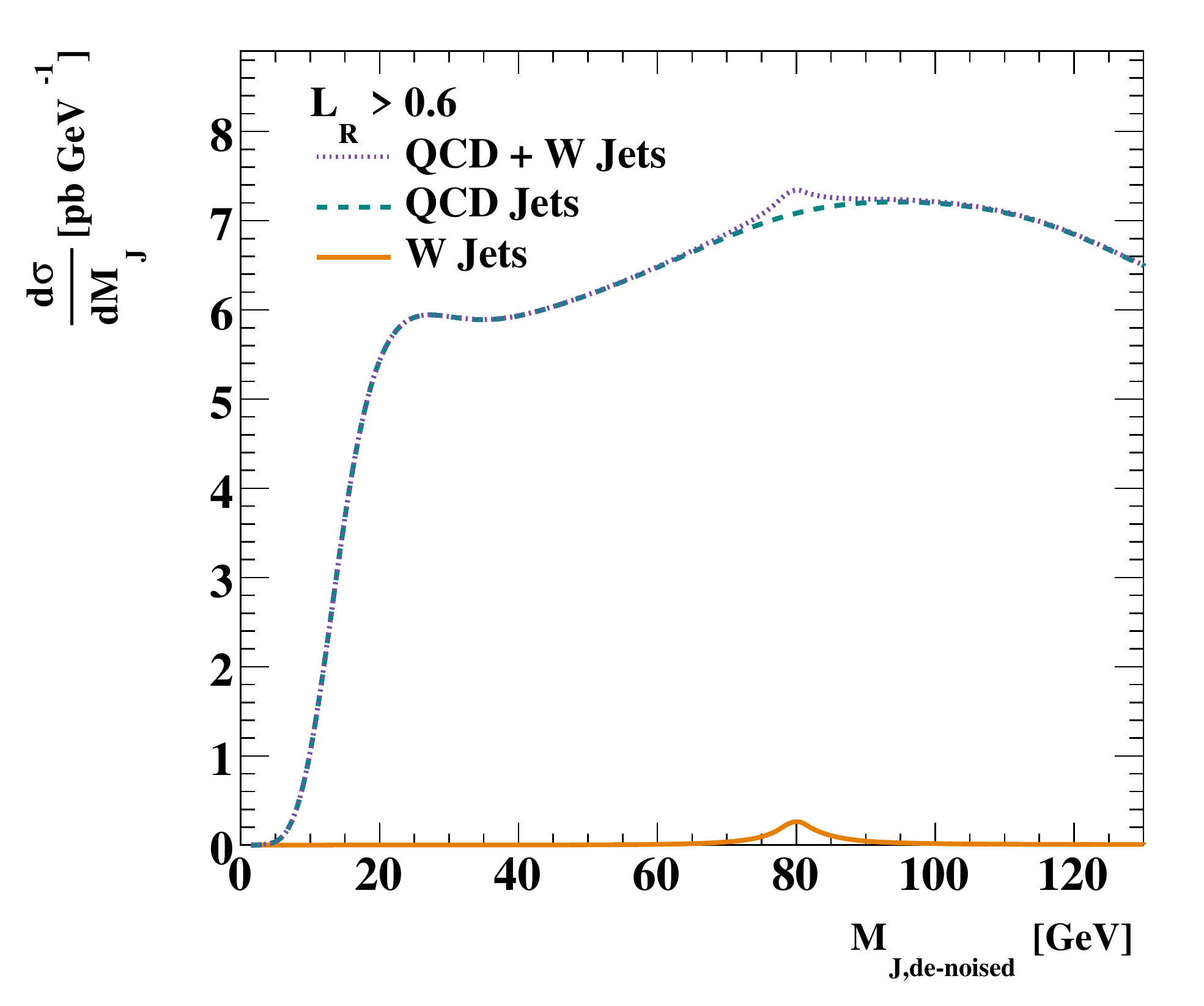}
\put(83, 25){a)}
\end{overpic}
%\begin{overpic}[width=0.7\columnwidth, angle=270]{massSelected_09}
\begin{overpic}[width=0.7\columnwidth]{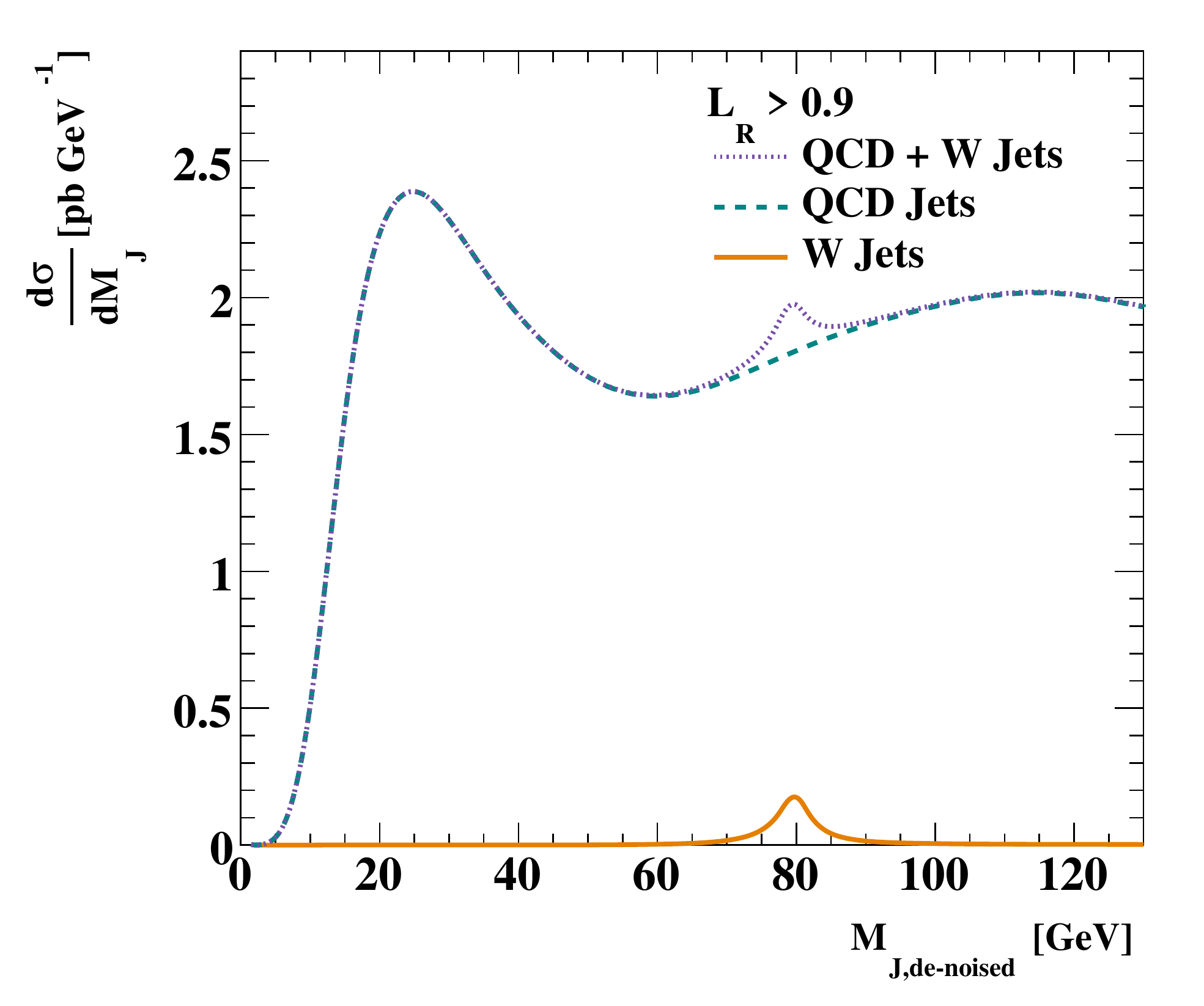}
\put(83, 25){b)}
\end{overpic}
\caption[Masses of jets selected by wavelet analysis]{Mass distribution of de-noised jets after selection by wavelet analysis.  A loose likelihood selection of $L_{\mathrm{R}} > 0.6$ that preserves most signal jets is shown in a), while a tight selection of $L_{\mathrm{R}} > 0.9$ that improves the signal to background ratio at the expense of the total number of signal jets is shown in b). }\label{fig:selectedMass}
\end{center}
\end{figure}

Figure \ref{fig:selectedMass} shows the de-noised jet mass distribution for QCD- and W-jets, together with their combination, when selecting jets that satisfy $L_{R} > \mathrm{0.6}$ and $L_{R} > \mathrm{0.9}$.  The process of wavelet de-noising followed by jet recognition hugely increases the signal to background ratio of W-jets to QCD-jets; the differential QCD-jet cross section in the region of the W-mass is around 55~$\mathrm{pb}^{-1}\mathrm{GeV}^{-1}$ (figure \ref{fig:filteredMass}b), and is reduced to around 1.8~$\mathrm{pb}^{-1}\mathrm{GeV}^{-1}$ by wavelet analysis with $L_{R}>\mathrm{0.9}$, a reduction of 30:1.  The signal W-jet mass distribution is on the other hand enhanced, although the $L_{R}$ selection requirement does reduce the signal peak in figure \ref{fig:selectedMass}b compared to figure \ref{fig:filteredMass}a.  Selecting jets with even larger values of $L_{\mathrm{R}}$ could further improve the signal to background ratio, though at the expense of the total number of signal events.

The shape of the QCD background distribution in figure \ref{fig:selectedMass} clearly depends on the value of the $L_{\mathrm{R}}$ selection criterion.  Changing the jet selection to require larger values of $L_{\mathrm{R}}$ enhances the bump in the QCD $M_J$ distribution at between 20 to 30~GeV.  This change in the shape of the distribution occurs because there are different contributions to the QCD-jets from Feynman diagrams involving both quarks and gluons.  While it is a fallacy to describe a single jet as either a quark or a gluon, the two (at a minimum) distinct populations of jets are a reflection of this underlying physics.  That wavelet analysis is sensitive to these details of QCD production is in itself very interesting, and may open up new tests of QCD.  Further, a more complete study may be able to separate these different QCD-jet populations, and by doing so further enhance the separation of signal W-jets.

\section{Effect of Multiple Overlaid Collisions}

Coping with the effects of multiple overlaid hadron collisions (known as ``pile up'') would seem to be an obvious application for performing analysis in the frequency domain.  Pile up adds a carpet of radiation throughout the detector, which adds to the momentum and mass of the jets, as well as creating jets that would not otherwise be present.  The overlaid pile up events are usually softer than the hard collision of interest, with few or no high \pT jets of their own.  Some approaches to pile up removal include jet area subtraction, which subtracts from jets a momentum contribution based on the average detector activity multiplied by the jet area, and jet substructure pruning techniques, which remove small radius low \pT jets from large radius high \pT jets.

These two approaches to pile up removal represent the far extremes of the frequency domain.  Jet area subtraction removes very wide angle contributions, arising from the average activity across the entire detector.  Jet pruning removes small localised contributions to large jets.  An appropriate wavelet analysis on the other hand offers the opportunity to remove both the large scale contributions from pile up, the small scale fluctuations that occur on top of the local background, and all angular scales in between.  

In order to perform studies with pile up, a further sample of Monte Carlo events was generated.  Pythia 8 \cite{Sjostrand:2007gs} tune A2 \cite{ATLAS:2012uec} with the CTEQ6L1 PDF \cite{Pumplin:2002vw} was used to generate a sample of 44 million soft inelastic events (including diffractive processes) at a centre of mass collision energy of 8~TeV.   These soft QCD events were combined with the sample of W bosons using the PileMC package \cite{pilemc}, which overlays a random number of pile up collisions on top of a signal sample.  The number of pile up collisions was distributed according to the Poisson distribution with a mean of 20.

De-noising $\mathcal{O}$(20) overlaid collisions differs from de-noising a single collision because the individual collisions are not correlated.  The noise threshold, $t\left(m_y, m_\phi\right)$ at wavelet levels $m_y$ and $m_\phi$ was therefore chosen according to equation \ref{eqn:threshold}

\begin{equation}
t\left(m_y, m_\phi \right) = \left(1 + N_{\mathrm{pileup}} \times\frac{6.4}{2^{m_y}\times2^{m_\phi}} \right)    \mathrm{GeV}\label{eqn:threshold}
\end{equation}
where $N_{\mathrm{pileup}}$ is the number of overlaid pile up collisions in the event being analysed.  Equation \ref{eqn:threshold} was motivated as follows:  each soft inelastic collision contributes around 1~GeV of activity per unit of rapidity, or $N_{\mathrm{pileup}}\times6.4$~GeV in total within $\left|y\right|<$3.2 for $N_{\mathrm{pileup}}$ events.  Each wavelet basis function at wavelet levels $m_y$ and $m_\phi$ is localised to an area of approximately $2\pi / 2^{m_\phi} \times 6.4 / 2^{m_y}$, which is a fraction $1 / (2^{m_y}\times 2^{m_\phi})$ of the total detector area available.  A rough estimate of the average pile up contribution to the wavelet coefficient at levels $m_y$ and $m_\phi$ is therefore 1~GeV per unit of rapidity, multiplied by the number of pile up collisions and the fraction of the total area to which the wavelet basis function is localised.  To this is added a 1~GeV pedestal, which is the noise threshold from section \ref{sec:denoise}, and accounts for soft contributions present in the hard signal collision.

While this choice of threshold has not been fully optimised, it demonstrates an important feature of the wavelet analysis; the threshold for very wide-angle contributions can be made much higher than that for small angle contributions.  The motivation for this is that the wide-angle coefficients receive contributions from all of the overlaid collisions together, whereas the small-angle coefficients are less likely to receive contributions from multiple collisions, but simply represent local fluctuations arising from a single collision.  The power of the wavelet analysis allows these different contributions, the detector-wide average and the local fluctuations, to be extracted and treated differently.  An in depth study of soft physics in the wavelet domain would allow for a better understanding of the most appropriate threshold as a function of angular scale.

\begin{figure}
\begin{center}
\includegraphics[width=0.8\columnwidth]{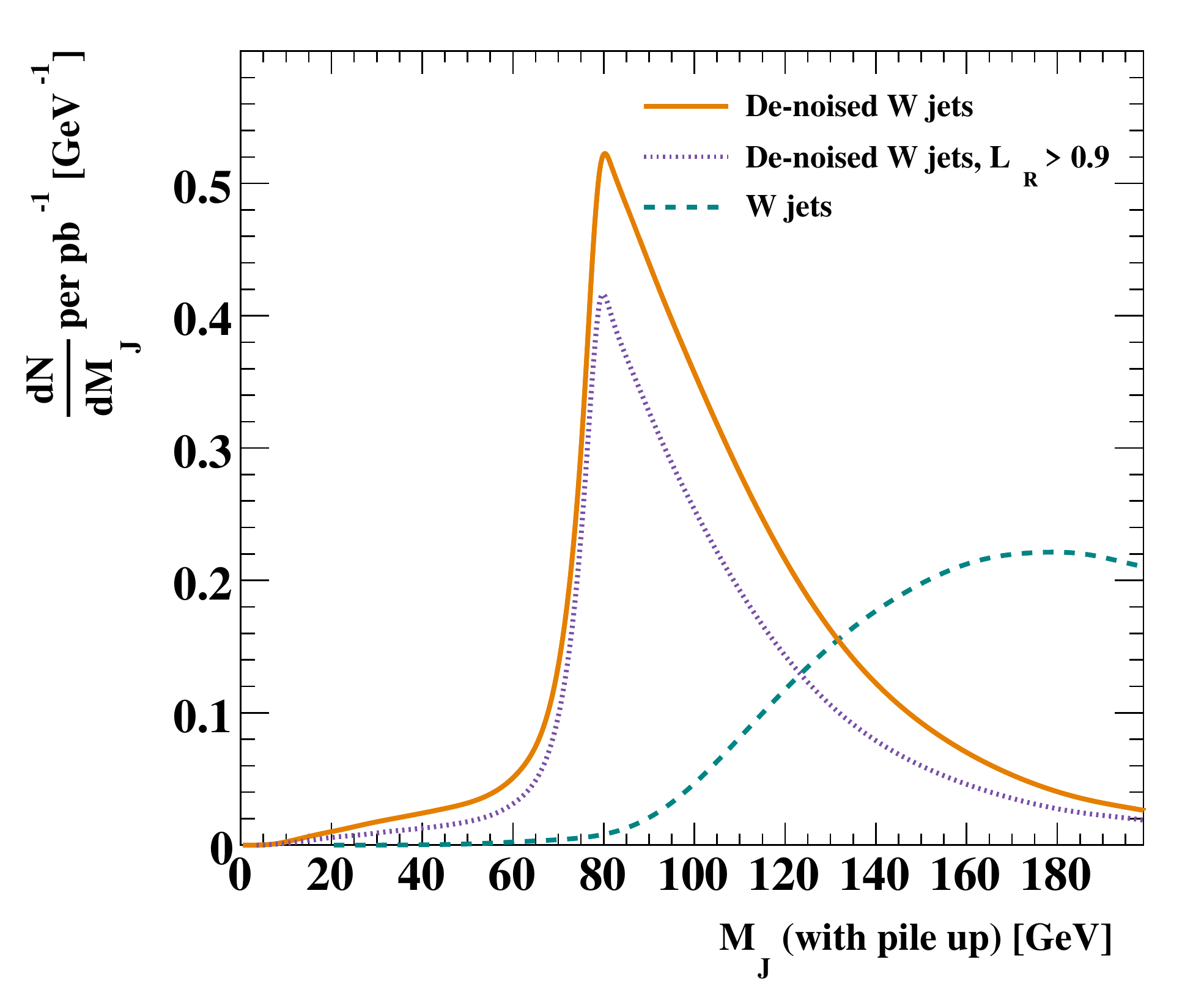}
\caption[Jet de-noising with pile up]{Effect of wavelet de-noising and selection on W-jet mass in the presence of pile up.  The dashed teal curve shows the matched jet mass distribution when an average of twenty soft inelastic collisions is overlaid.  The solid orange curve shows W-jet masses from the same events after wavelet de-noising to eliminate pile up.  The dotted purple curve shows the masses of those jets that pass the wavelet jet selection with a likelihood $L_{R}>0.9$.}\label{fig:pileupMass}
\end{center}
\end{figure}

Figure \ref{fig:pileupMass} shows the jet mass for the signal sample of W bosons after pile up has been overlaid.  Prior to any wavelet analysis, the peak in the distribution is extremely broad, and has shifted to around 170~GeV.  After wavelet processing to remove those wavelet contributions below the threshold in equation \ref{eqn:threshold}, there is a clear peak at the W-boson mass of 80~GeV.  Compared to the sample without pile up, this peak is in fact higher, but the pile up has distorted the shape of the peak, with a notable tail to higher jet mass.  Applying the likelihood-based jet selection does not affect either the shape or the height of the signal peak very much.

\begin{figure}
\begin{center}
\includegraphics[width=0.8\columnwidth]{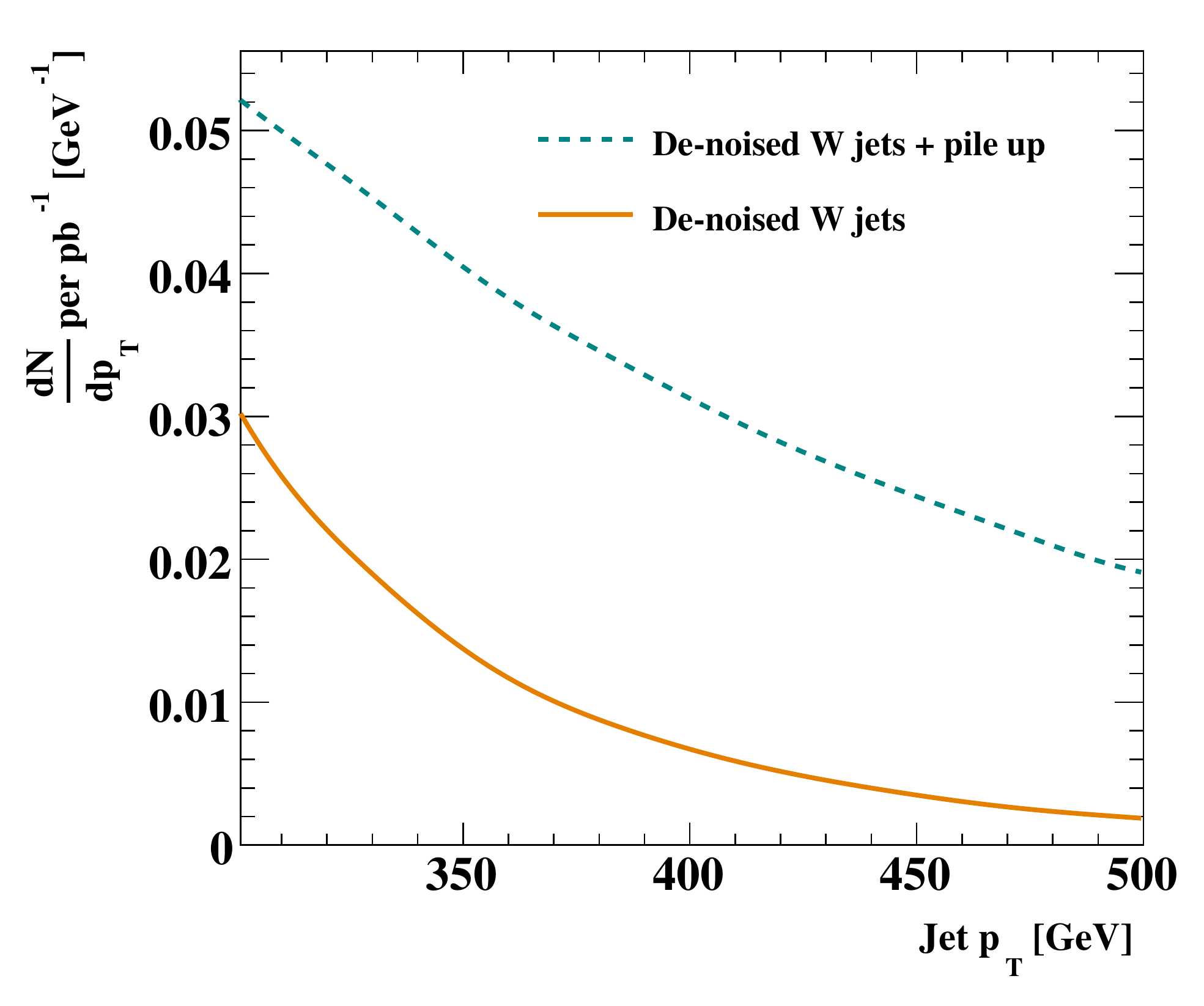}
\caption[Effect of pile up on de-noised jet \pT]{The effect of pile up on the de-noised jet \pT distribution.  The dashed teal curve shows the predicted number of jets when pile up is present, the solid orange shows the number in the absence of pile up.}\label{fig:pileupPT}
\end{center}
\end{figure}

The bias visible in the mass distribution may reflect a residual bias in the jet \pT distribution.  The jet \pT distribution for de-noised W-jets with and without pile up are shown  in figure \ref{fig:pileupPT}.  The jets with pile up have a harder \pT spectrum, which explains the greater height of the peak in the jet mass distribution with pile up.  The jet-mass will also be correlated with jet \pT, so the bias to higher \pT caused  by pile up will also favour higher jet masses.

Despite the remaining distortion in the shape of the mass distribution, wavelet analysis \emph{alone} has successfully retrieved a clear mass peak from a very broad mass distribution that reflects pile up conditions consistent with the first data taking runs of the Large Hadron Collider (LHC).  Further optimisation of the wavelet threshold, a better understanding of the \pT bias and the combined use of other pile up techniques (such as using vertex information from charged particle tracking detectors) offer plenty of scope for improving the behaviour in the presence of pile up beyond this initial study.

\section{Comments}

A large number of jet substructure techniques now exist, many of which have aspects that are similar to the wavelet analysis outlined in this note.  The jet energy flow approach of \cite{Berger:2002jt}, which predates mainstream jet substructure, is somewhat similar in that it convolutes the event energy flow with a smoothing function of variable radius.  Like the present wavelet analysis, that approach aims to avoid assigning an individual particle to a unique jet as a proxy for a hard parton.  However, despite introducing the concept of viewing an event over variable angular scales, that simpler smearing does not include the key ability to invert the event convolution, nor does it use orthogonal basis functions that are well organised on a dyadic grid with a clear scaling rule.  

The two point correlation approach of \cite{Jankowiak:2011qa} has some similarity with the Haar wavelet, which is a simple step function.  That approach also builds a top tagger by measuring the mass fraction as a function of correlation length.  Again, it is not as general as the wavelet analysis, nor does it include key concepts like de-noising.

Shower deconstruction, described in \cite{Soper:2011cr}, also shares some similarities with wavelet analysis.  In particular, the recognition of jets through their radiation patterns, and the ordered ``shower history'' concept is similar to the idea of using wavelets to probe the scale dependence of the shower.  However, where shower deconstruction requires specific signal and background models, wavelet analysis uses self-similarity and the different scale dependence of the signal compared to the background.  If the wavelet basis functions were transposed into a toy shower model and used to train shower decomposition, then the outcome would likely be similar to wavelet analysis, with the difference that wavelet analysis would be much simpler to implement.  How similar a wavelet shower model is to a real shower  is likely to determine the different performance of the two approaches.

Significantly, wavelet analysis is \emph{not} a jet substructure technique.  The ideas presented here are quite general and explain \emph{how} one could perform a wavelet analysis of hadron collisions, not what one might do with that analysis.  The key ideas are that of rasterisation, followed by wavelet decomposition and analysis, before re-composing the event with the modified wavelet components.  The concepts of de-noising, probing events over different angular scales and shower evolution are also introduced.  Jet substructure is a timely, popular and appropriate topic for the application of wavelet analysis, but the concept of using an event's angular scale dependence could equally also be used to search for new physics or understand QCD better.  Wavelet analysis is a hitherto largely ignored technique that could open provide an important new tool for studying hadron collisions.

\section*{Acknowledgements}

Thanks to Troels Petersen, who enthusiastically supported this idea and gave invaluable feedback.  Thanks also to Mario Campanelli, with whom the initial ideas for a harmonic analysis of QCD were developed, and Frank Krauss, who subsequently suggested that wavelets might make a good basis in which to perform such a study.  I also had several useful discussions with Lily Asquith, whose ``LHCSound'' project caused several conversations on the relevance of harmonic analysis to particle physics.  This work was funded by a grant from the Lundbeck foundation.

\appendix

\section{Example Using the Haar Wavelet}\label{apdx:example}

The Harr wavelet is the simplest basis wavelet, consisting of a step function that calculates the difference between adjacent values in a sequence of numbers that forms a signal.  While the Harr was not used for the wavelet analysis of hadron collisions, it is simple enough that it can be used for decomposition by hand, and as such forms a useful introduction to wavelets.

\begin{figure}
\begin{overpic}[width=\columnwidth]{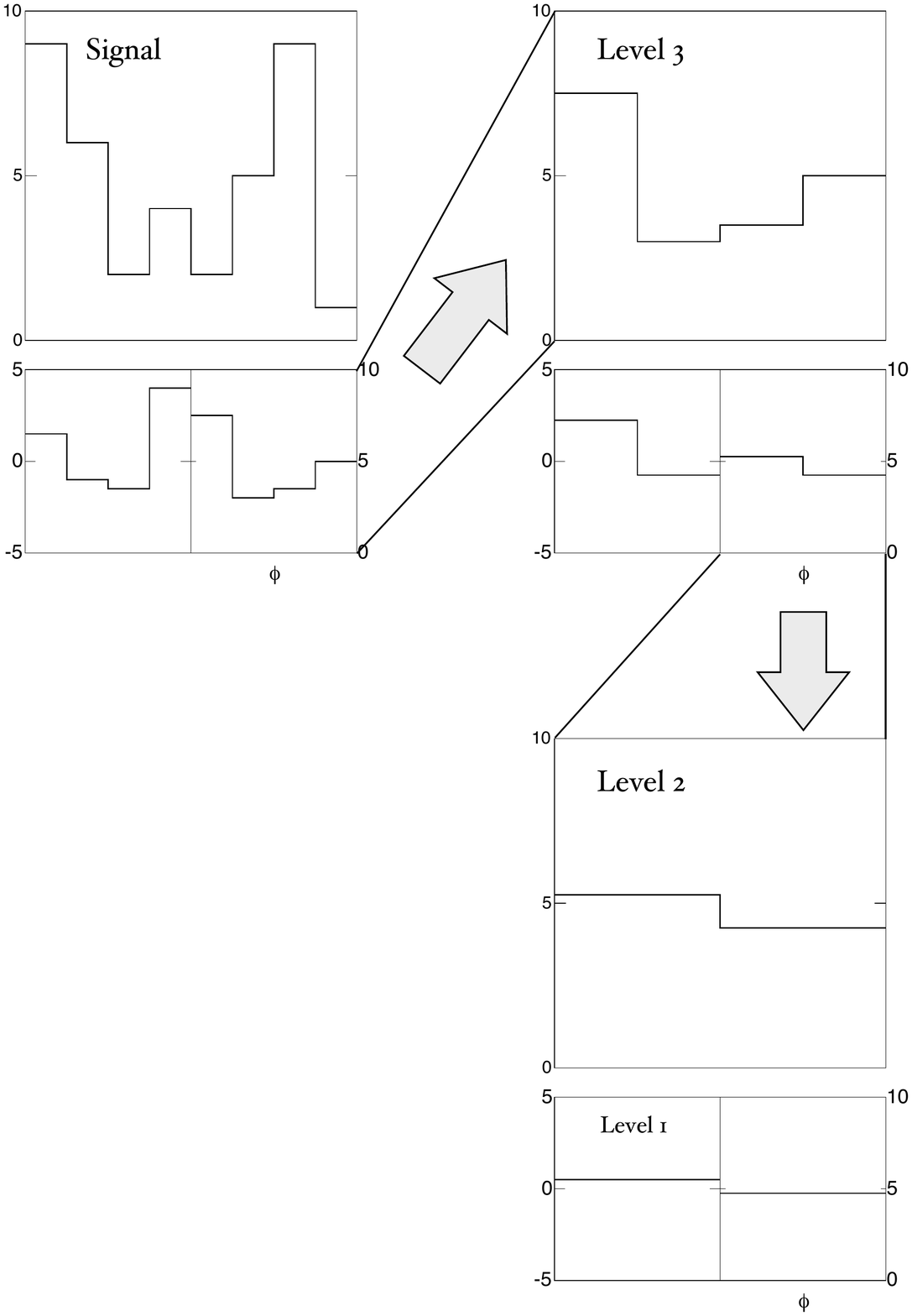}
\end{overpic}
\begin{flushleft}
\begin{minipage}{0.5\columnwidth}
\vspace{-13cm}\caption[]{The Harr decomposition of an 8-bin signal.  The input signal is in the top left, immediately below which are the level-1 wavelet coefficients (left) and the re-scaled signal (right).  The top right shows the signal after re-scaling at level-1, the two smaller panels below that show the level two coefficients (left) and the signal (right) after a second rescaling.  The bottom right shows the signal at the largest scaling factor (4, in this case), together with the large scale wavelet coefficient below it (left) and the average of the signal (right).  The signal can be fully reconstructed from the 7 wavelet coefficients together with the average value of the signal. }\label{fig:harr}
\end{minipage}
\end{flushleft}
\end{figure}

The re-scaling of the signal from one level to the next is performed by averaging adjacent values in the signal sequence.  The input signal must have a  length that is radix two.   The decomposition process is applied iteratively at each level of the decomposition, namely the high frequency part of the signal (the differences between adjacent values in the sequence) is separated from the low frequency part of the signal (the average of adjacent values).  The next level of the decomposition applies the same decomposition to the low frequency part of the previous level, the number of levels being limited by the length of the input sequence.  Most wavelet basis functions are more complicated than the Harr wavelet, and the re-scaling between levels is also more complicated than a simple average between adjacent values, but this basic iterative principle is still used.  Figure \ref{fig:harr} illustrates the iterative decomposition of the signal.

\end{document}